\documentstyle[11pt]{article}

\begin{document}

\begin{center}
{\LARGE \bf Interplay between Heavy Fermions and Crystal Field 
Excitation in Kondo Lattices. Low-Temperature Thermodynamics  
and Inelastic Neutron Scattering Spectra of CeNiSn } 
\mbox{} \\ \mbox{} \\ \mbox{} \\ 
{\Large\bf Yu. Kagan, K.A. Kikoin and A.S. Mishchenko}\\ \mbox{} \\ \mbox{} \\ 
{\Large\it Kurchatov Institute, Moscow 123182, Russia}
\end{center}
\vspace{8mm}
\begin{abstract}
{\large The microscopic theory of interaction between the heavy fermions and the 
crystal field excitations in Kondo lattices is presented. It is shown  
that the heavy-fermion spectrum scaled by the Kondo temperature $T_K$ can 
be modified by the crystal field excitations with the energy $\Delta_{CF}$ 
provided the inequality $\Delta_{CF}<T_K$ is realized. On the base of general 
description of excitation spectrum the detailed qualitative and quantitative 
explanation of anisotropic inelastic neutron scattering spectra  
and low-temperature specific heat of orthorhombic CeNiSn is given. 
The theory resolves the apparent contradiction 
between the metallic conductivity and the gap-wise behavior of thermodynamic 
properties and spin response of CeNiSn at low temperatures.  }
\end{abstract}
\mbox{} \\ \mbox{} 
\mbox{} \\ \mbox{} \\ 
{\bf Contact address:} \\ \mbox{} \\ 
{\large
A.S.Mishchenko \\
RRC Kurchatov Institute, Moscow 123182, Russia\\
tel: (7)(095)1969148; \\ fax: (7)(095)1965973;
\\ e-mail: andry@kurm.polyn.kiae.su}

\newpage

\section{Introduction}
Although the heavy fermion (HF) behavior of the rare earth- and actinide-based 
intermetallides is related to the localized electrons 
in the 4f- and 5f-shells of Ce and U, the influence of the crystal field (CF) 
splitting of the f-levels is inessential, 
as a rule, for the low-energy excitations in 
these systems, since the inequality $\Delta_{CF}\gg T^*$ is valid for  
most of these compounds \cite{Fulde95}. Here $T^*\sim T_K$ is the 
temperature which characterizes the energy scale of the HF spectrum.  
The only role of this interaction is to form the ground state Kramers 
doublet which is the eventual source of  the low-energy excitations 
with extremely high density of states. 
However, the picture changes radically when this inequality is violated. 
One can expect that in case of $\Delta_{CF} < T^*$ the inevitable 
interplay between the local CF excitations and itinerant HFs should result 
in appearance of the fine structure in the low-energy spectrum which should
be characterized by at least two additional energy parameters, 
i.e., the position of the CF level and the magnitude of intermixing.    
This interplay should result in remarkable changes in the spectral and 
thermodynamic properties of the system in the temperature interval 
$0< T < \Delta_{CF}$, although it is clear that the characteristic HF 
behavior of the system should persist at $T\rightarrow 0$: 
the Fermi liquid-like temperature dependences for the specific heat, 
$C(T)\sim T$, electrical resistivity, $\rho(T)-\rho(0)\sim T^2$, NMR 
relaxation rate, $T_1T$= const, etc., should be observed,   
although the coefficients in these laws should be sensitive not only to 
$T^*$ but also to these new parameters. On the other hand, the 
hybridization of the HF and CF excitation can result, as it frequently occurs, 
in appearance of the psedogap in the energy spectrum, although this 
pseudogap should be characterized mainly 
by the crystal field and mixing parameters.  

The idea of applying these considerations to description of unusual  
properties of CeNiSn was proposed in \cite{KKP93}. It was shown 
that one of the necessary preconditions for realization of this mechanism 
is the specific picture of CF level splitting: it was supposed that two 
lowest Kramers doublet of Ce(f$^1$) configuration are separated by a 
small energy gap 
\begin{equation}
\Delta_{CF}<T^*
\label{0.1}
\end{equation}
while the distance between the ground state  and the 
second excited level is much larger, $\Delta_{CF}'\gg \Delta_{CF}$.
The characteristic HF energy $T^*$ is estimated as 40 to 80 K from 
various experimental data for CeNiSn \cite{Taka93}. 
Later on, 
the indirect measurements by Alekseev et al \cite{Aleks94}  
confirmed the assumption (\ref{0.1}). 
In that paper the crystal field was measured on the Nd ion in
Nd$_{0.3}$La$_{0.7}$Ni crystal which is equivalent to CeNiSn 
from the point of view of RE crystalline 
environment and lattice spacings. 
The crystal field acting at the RE site was restored from 
measured CF splitting of Nd f-level and then recalculated for the case of Ce 
ion with the same parameters of equivalent CF Hamiltonian in approximation of 
purely electrostatic field. The CF splitting energies were estimated as 
$\Delta_{CF}\approx 4.4$ meV and $\Delta_{CF}'\approx 14.0$ meV which is 
consistent with the assumption of ref. \cite{KKP93} 
and the inequality (\ref{0.1}). 

Another experimental result crucial for the above general consideration  was
the observation of Fermi-liquid behavior at low enough temperature in good 
quality samples of CeNiSn and CeRhSb. It was found that these samples 
demonstrate the metallic 
$T^2$-dependence of electrical resistivity at $T<7$K \cite{Taka96}, the 
NMR relaxation rate returns to Korringa law at $T<1$K \cite{Nakam95}, and 
the limiting value of the Sommerfeld 
coefficient $\gamma$ in the linear-T term of $C(T)$ is 
$\gamma=40$ mJ/K$^2$mol at 0.03 K \cite{Taka96}. This value is noticeably  
lower than that predicted by the high-temperature estimations of $T^*$. 
These results seemed to be striking, since the semiconductor-like rise 
of the resistivity and $T^3$ dependence of $1/T_1$ was observed 
earlier on the less perfect samples \cite{Taka93}.

The most challenging experimental observations on CeNiSn are the inelastic 
magnetic neutron scattering spectra which demonstrate extremely complicated 
$({\bf Q},\omega)$-dependent structures with quasi gap behavior 
for some directions of ${\bf Q}$ \cite{Mason92,Kadow94,Sato95,Kambe96}. 
All these features disappear completely at T=25 K. This picture is 
inexplicable in terms of standard notions of intersite spin fluctuations. 
At the same time, these data are highly informative from the point of 
view of the structure of low energy spin excitations, so the adequate 
interpretation of the neutron scattering data is the most effective 
criterion for selecting out of various theoretical interpretations.

In the present paper the spectral properties of CeNiSn are analyzed on 
the base of the ideas formulated in \cite{KKP93} and applied for the 
quantitative explanation of low temperature thermodynamics in 
\cite{KKP93,KVBT94}. We apply these ideas to the spin liquid which is 
formed in Kondo lattices, provided the magnetic order does not appear 
at low temperature. We describe this spin liquid in terms of resonating 
valence bond (RVB) excitations similar to those introduced by P. Anderson 
and used later in description of nearly magnetic state in 2D Cu-O planes 
of high-T$_c$ materials \cite{Bask87}. In this case the characteristic 
properties of HF liquid are determined by the subsystem of spin 
excitations (spinons) which obey Fermi statistics at low temperature. 
The stabilization mechanism for the RVB state in 3D Kondo lattices was 
offered in \cite{KKM94}. 
We modify the theory of spin liquid for the case of $\Delta_{CF}<T_K$, 
calculate the spectrum of RVB excitations for a special case of 
orthorhombic CeNiSn lattice and give a detailed comparison of calculated 
neutron scattering cross section with the experimental results of 
\cite{Sato95,Kambe96}. The quasi 2D character of spin excitations 
discovered in these experiments is essentially used in our calculations. 
Quite reasonable agreement between the theoretical curves and experimental
constant-${\bf Q}$ and constant-$E$ scans seems encouraging. To make 
an additional test of adequacy of the theory, we calculate also the 
low-temperature specific heat of CeNiSn within the same set of model 
parameters.  
\section{Degenerate spin liquid in a crystal field}

The theory of HF systems starts with  
the Anderson lattice Hamiltonian. In a special case of 
Ce$^{(3+)}(f^1)$ ions in a crystal field this Hamiltonian is usually 
written as 
\begin{equation}
H = H_{f} + H_{c} + H_{cf} + H' 
\label{1.1}
\end{equation}
Here $H_{f}$ describes the Ce ions in a crystal field,  
\begin{equation}
H_{f} = 
\sum_{{\bf i}, \Lambda} E_{\Gamma} 
f^{\dagger}_{{\bf i}\Lambda}f_{{\bf i}\Lambda} +
\frac{U}{2}\sum_{{\bf i},\Lambda \neq \Lambda'}  
f^{\dagger}_{{\bf i}\Lambda}f_{{\bf i}\Lambda}
 f^{\dagger}_{{\bf i}\Lambda'}f_{{\bf i}\Lambda'}\;,
\label{1.2}
\end{equation}
$f^{\dagger}_{{\bf i}\Lambda}$ is the creation operator of the f-electron 
in a state $\Lambda=\Gamma\nu$ in a site ${\bf i}$, where $\nu$ is the row 
of the irreducible representation $\Gamma$ of the crystal point group,
$E_{\Gamma} $ denotes the energy of the 
f-level in the electrostatic crystal field, and $U$ stands for 
the Coulomb repulsion between the f-electrons in the same site. 
The f-ions are immersed in a Fermi-sea of 
conduction electrons which are described by the Hamiltonian
\begin{equation}
H_{c} = \sum_{k,\Lambda} \epsilon_{k\Gamma} 
c^{\dagger}_{k\Lambda} c_{k\Lambda}. 
\label{1.3}
\end{equation}
The third term in the Hamiltonian (\ref{1.1}) describes the hybridization 
between f- and conduction electrons,
\begin{equation}
H_{cf} = \sum_{kk'} \sum_{{\bf i}} \sum_{\Lambda} 
\left(
V_{k\Lambda}^{\bf i}
f^{\dagger}_{{\bf i}\Lambda} c_{k\Lambda}+ H.c.
\right)
\label{1.4}
\end{equation}
The so called Coqblin-Cornut (CC) approximation \cite{Coqb69,Corn72}
which represents the Bloch functions by their partial waves 
$c^{\dagger}_{k\Lambda}$, and takes into account only the diagonal 
in $\Lambda$ hybridization matrix elements 
$V_{k\Lambda}^{\bf i}=\langle{\bf k}\Lambda|V|{\bf i}\Lambda\rangle$ 
is used. 
All the one-electron processes beyond this approximation are collected in the 
last term $H'$ in the Hamiltonian (\ref{1.1}). 

It is widely believed that the s=1/2 approximation for the total moment $j$ 
of Ce(f$^1$) ion works well in the Anderson lattice Hamiltonian 
because the CF splitting of the sextet j=5/2 
usually results in a simple Kramers doublet ground state, and the CF 
splitting exceeds essentially the characteristic interaction energy $T_K$. 
This procedure seems to be reasonable when one deals with the ground 
state and the low-energy excitations with characteristic frequencies 
$\hbar\omega <\Delta_{CF}$. Since our task is inclusion of 
CF excitations in the general picture of spin liquid behavior, we should 
begin with  generalization of s=1/2 theory for the case of realistic crystal 
field scheme. 

The Kondo lattice deals with the well localized f-electrons for which the 
inequality $V_{{\bf k}\Lambda}\ll\epsilon_F-E_f$, where 
$E_f$ is the energy level of free Ce(f$^1$) ion, is assumed to be valid, and 
the Hubbard parameter $U$ is taken large enough to suppress the doubly 
occupied f-states of this ion. 
In this case the canonical transformation eliminating the hybridization 
interaction from the Hamiltonian (\ref{1.1}) can be done (see \cite{Corn72}), 
and the effective Hamiltonian projected to the subspace of homopolar states 
with $n_f=1$  can be written as 
\begin{equation}
H^{(CC)} = H_{f} + H_{c} + H_{ex} + H_{h} 
\label{1.5}
\end{equation}
where 
\begin{equation}
H_{ex} =  \sum_{kk'} 
\sum_{\bf i} \sum_{\Lambda, \Lambda'}
J_{\bf i}^{\Lambda\Lambda'}(k,k') 
f^{\dagger}_{{\bf i}\Lambda} f_{{\bf i}\Lambda'} 
c^{+}_{k'\Lambda'} c_{k\Lambda}
\label{1.6}
\end{equation}
describes the effective sf-exchange interaction, and 
\begin{equation}
H_{h} = - \sum_{k} 
\sum_{\bf i} \sum_{\Lambda}
J_{\bf i}^{\Lambda\Lambda}(k,k) 
f^{\dagger}_{{\bf i}\Lambda} f_{{\bf i}\Lambda}\;. 
\label{1.61}
\end{equation} 
corresponds to effective covalent contribution to the one-site CF 
splitting due to virtual sf-transitions. 
Here  
\begin{equation}
J_{\bf i}^{\Lambda\Lambda'}(k,k')=
\frac{V_{k\Lambda}^{{\bf i}*}V_{k'\Lambda'}^{\bf i}}
{\epsilon_k-E_f} \;.
\label{1.7}
\end{equation}
We assume for the sake of simplicity that the conduction electrons  
are degenerate in $\Lambda$ near the electron Fermi surface and neglect   
the CF splitting of the f-level in the denominator of 
effective sf-exchange integral. 
\par
We are interested in the case of $n_f=1$ when the charge fluctuations 
in the f-channel are completely suppressed.  Then, according to the scenario 
of \cite{KKM94,KKP92} the neutral spin liquid should arise at some temperature 
$T^*\sim T_K$ instead of antiferromagnetic or Kondo-singlet state offered 
by the standard Doniach's dichotomy. And, as a result,  
the spin liquid excitations together with the low-energy 
electrons turn out to be responsible 
for the low-temperature thermodynamics of the Kondo lattice. 
Now we incorporate the CF excitations in this picture. 
The "fast" electrons with characteristic energies 
$\hbar\omega>T_K$ are responsible for formation of the spin-spin correlation 
at $T>T_K$ 
(the Kondo temperature is defined as $T_K\approx \epsilon_f\exp(-1/2\alpha)$, 
where $\alpha=J_{sf}{\cal N}(\epsilon_F$), 
$J_{sf}=|V_{k_F}|^2/(\epsilon_F-E_f)$, and
${\cal N}(\epsilon_F)$ is the electron density 
of states at Fermi level).
These fast electrons can be integrated out, and the effective Hamiltonian 
for the spin degrees of freedom can be written in the following form.
\begin{equation}
H^s =H_f + H_h^{\prime} +  H_{RKKY}^{(g)}+H_{RKKY}^{(nd)}\;, 
\label{1.900}
\end{equation}
\begin{equation}
H_{RKKY}^{(g)} = 
\sum_{\bf ii'}^{{\bf i} \ne {\bf i}'} \sum_{\nu\nu'} 
I^{GG}_{\bf ii'} 
f^{\dagger}_{{\bf i}G\nu} f_{{\bf i}G\nu'} 
f^{\dagger}_{{\bf i'}G\nu'} f_{{\bf i'}G\nu} \;,
\label{1.9}
\end{equation}
\begin{equation}
H_{RKKY}^{(nd)} = 
\sum_{\bf ii'}^{{\bf i} \ne {\bf i}'} \sum_{E\nu\nu'}
I^{GE}_{\bf ii'} 
f^{\dagger}_{{\bf i}G\nu} f_{{\bf i}E\nu'} 
f^{\dagger}_{{\bf i'}E\nu'} f_{{\bf i'}G\nu} 
\label{1.18}
\end{equation}
Here the notations $\Lambda=G\nu$ and $\Lambda=E\nu$ are used for the 
ground state and excited states, respectively, 
$H_h^{\prime}$ includes the scattering correction to the CF level 
renormalization due to the interaction $H_{ex}$,  
\begin{equation}
I^{GG}_{\bf ii'}\sim 
\alpha^2\Phi_G({\bf k}_F,{\bf R}_i-{\bf R}_{i'})
B_G(\theta, \phi)K(T)
\label{1.10}
\end{equation} 
is the indirect exchange interaction which contains oscillating RKKY function 
$\Phi_G({\bf k}_F,{\bf R}_i-{\bf R}_{i'})$, aniso\-tropy factor 
$B_G(\theta, \phi)$ 
($\theta$ is the angle between the $z$-axis which is assumed to be the 
axis of magnetic quantization, and the line connecting the ions 
${\bf R}_i$ and ${\bf R}_{i'}$ and $\phi$ is the angle of rotation about 
$z$-axis) (see Appendix), 
and additional enhancement factor $K(T)$  due to the 
high-temperature one-site Kondo scattering \cite{KKM94}.  
$I^{GE}_{\bf ii'}$ is the integral of the same type as 
that in eq. (\ref{1.10}), with its own factors $B_{GE}$ and $\Phi_{GE}$, 
and $K_{GE}(T)=1$.   

The uniform spin liquid state of RVB type in a standard 
s=1/2 Hesenberg model 
is described in terms of correlators \cite{{Bask87}},
\begin{equation}
\Delta=\sum_{\sigma}
\langle f^{\dagger}_{{\bf i}\sigma}f_{{\bf i'}\sigma}\rangle,\; i \ne i'  .
\label{1.800}
\end{equation}
It was shown in \cite{KKM94} that the weak-coupling Kondo interaction at 
$T>T_K$ favors the stabilization of RVB state against antiferromagnetic 
ordering because the Kondo scattering screens the local moment leaving 
intact the singlet RVB correlators (\ref{1.800}). 
This stabilization "quenches" 
the Kondo scattering at temperatures $T\sim T^*>T_K$ and allows 
the description of spin liquid in terms of RVB correlation functions 
at $T<T_K$, since $T_K$ is no more a singular point of perturbation 
theory. Having in mind this stabilization mechanism we use the mean-field 
approximation for the  uniform RVB state even at $T<T_K$. 
Although the interaction with "slow" conduction electrons and the influence 
of spin and gauge fluctuations can modify the properties of spin-fermions, 
we use this mean-field description as the basic approximation for studying 
the system with the CF excitations involved.  

Thus, the correlator similar to (\ref{1.800})
can be introduced for the ground state doublet $G\nu$:
\begin{equation}
\Delta^G= \sum_{\nu}
\langle f^{\dagger}_{{\bf i}G\nu}f_{{\bf i'}G\nu}\rangle, \; i \ne i' .
\label{1.8}
\end{equation} 
The spin-fermion excitations which are described by the Hamiltonian 
(\ref{1.900}) are constrained by the condition  
\begin{equation}
\sum_{\Lambda} f^{\dagger}_{{\bf i}\Lambda}f_{{\bf i}\Lambda}=1\;. 
\label{1.16}
\end{equation} 

Important contributions to the interplay between the low-energy spinon 
excitations and the one-site CF excitations
are connected with the interactions $H'$ which 
appear in the Hamiltonian (\ref{1.1}) beyond CC  
approximation. The origin of these interactions is the hybridization 
$\bar{V}_{k\Lambda}^{{\bf i}\Lambda'}=
\langle {\bf i}\Lambda|V'|k\Lambda'\rangle$ where $V'$ is the 
component of the crystal field which has the symmetry {\it lower} than that 
diagonalizing the energy terms $E_{\Gamma}$. 
When the integrals $\bar{V}_{k\Lambda}^{{\bf i}\Lambda'}$
are included in the canonical transformation, 
additional effective exchange terms appear: 
\begin{equation}
H_{ex}' =  \sum_{kk'} 
\sum_{{\bf i},E,\nu\nu'\nu''}
\left[
\frac{\bar{V}_{kG}^{{\bf i}E*}V_{k'G}^{\bf i}}
{\epsilon_k-E_f}
f^{\dagger}_{{\bf i}E\nu} f_{{\bf i}G\nu'} 
c^{+}_{k'G\nu'} c_{kG\nu''}+
\frac{V_{kG}^{{\bf i}*}\bar{V}_{k'E}^{{\bf i}G}}
{\epsilon_k-E_f}
f^{\dagger}_{{\bf i}G\nu} f_{{\bf i}G\nu'}
c^{+}_{k'E\nu''} c_{kG\nu} 
+ H.c.
\right] 
\label{1.60}
\end{equation}
These terms being combined with CC exchange term (\ref{1.6})
gives additional contributions to the intersite effective 
exchange, and the first order terms in $\bar{V}$ have the following form, 
\begin{equation}
\bar{H}_{RKKY}^{nd} = 
\sum_{\bf ii'}\sum_{E\nu\nu'\nu''}\left[
\bar{I}^{GE}_{\bf ii'} 
f^{\dagger}_{{\bf i}G\nu'} f_{{\bf i}E\nu} 
f^{\dagger}_{{\bf i'}G\nu} f_{{\bf i'}G\nu''} + H.c \right] \;.
\label{1.182}
\end{equation}

Here $\bar{I}^{GE}_{\bf ii'}$ is the integral of the same type as 
that in eq. (\ref{1.10}), but the corresponding coupling constant 
$\alpha'$ is proportional to $V^3\bar{V}$ and the factors $\bar{B}_{GE}$ and 
$\bar{K}(T)$ differ from those in (\ref{1.10}). 

Then, introducing the correlator $\Delta^G$ (\ref{1.8}), 
we obtain the mean-field  Hamiltonian for the spin-fermion spectrum
\footnote{The non-diagonal term (\ref{1.18}) gives no contribution to the 
uniform mean-field RVB pairing and describes only the fluctuation 
corrections to the mean-field solutions. These corrections are beyond the 
framework of the present paper}: 
\begin{equation}
H_{MF} = \widetilde{E}_G
\sum_{{\bf i},\nu} 
f^{\dagger}_{{\bf i}G\nu} f_{{\bf i}G\nu}+
\sum_{\bf ii'}^{{\bf i} \ne {\bf i}'} \sum_{\nu} 
{\cal T}_{\bf ii'}
f^{\dagger}_{{\bf i}G\nu} f_{{\bf i}'G\nu} + \delta H_{MF}\;,
\label{1.11}
\end{equation}
where
\begin{equation}
\delta H_{MF} = \widetilde{\Delta}_{CF}
\sum_{{\bf i}\nu}
f^{\dagger}_{{\bf i}E\nu} f_{{\bf i}E\nu}+
\sum_{{\bf i}\nu\nu'}
\left[{\cal B}^{GE}_{\bf ii}f^{\dagger}_{{\bf i}G\nu} f_{{\bf i}E\nu'}
+H.c.\right]
+\sum_{\bf ii'}^{{\bf i} \ne {\bf i}'}
\sum_{\nu\nu'}
\left[ 
{\cal B}_{\bf ii'}^{GE}
f^{\dagger}_{{\bf i}G\nu} f_{{\bf i'}E\nu'} + H.c.
\right]
\label{1.19}
\end{equation}
We use the bare ground state CF level as a reference point, $E_G=0$, then 
\begin{equation}
\widetilde{E}_{G} =
- N^{-1}\sum_{{\bf k},k>k_F} 
J^{GG}_{\bf i}({\bf k,k}) + 
\sum_{{\bf i}'}^{{\bf i}'\ne{\bf i}}  
I^{GG}_{{\bf ii}'} \equiv B_c^G + B_{ex}^G
\label{1.12}
\end{equation}
includes the covalent and indirect exchange contribution to CF shift
determined by the interactions (\ref{1.61}) and (\ref{1.9}), respectively,
and 
\begin{equation}
\widetilde{\Delta}_{CF}= \Delta_{CF} - N^{-1}\sum_{{\bf k},k>k_F} 
J^{EE}_{\bf i}({\bf k,k}) \equiv B_c^E 
\label{1.120} 
\end{equation}
also includes the one-site covalent corrections. 
The coupling constant in the MF Hamiltonian is given 
by the following equation, 
\begin{equation}
{\cal T}_{\bf ii'} = \sum_{\nu}
I^{GG}_{\bf ii'} 
\left\langle
f^{\dagger}_{{\bf i}G\nu} f_{{\bf i}'G\nu} 
\right\rangle=
I^{GG}_{\bf ii'}\Delta^G
\;.
\label{1.13}
\end{equation}
Only the sites connected by antiferromagnetic RKKY coupling contribute to 
the RVB correlation function.  

The one-site and intersite mixing constants are determined as
\begin{equation}
{\cal B}_{\bf ii}^{GE}=-\sum_{\bf k}
\frac{\bar{V}_{kG}^{{\bf i}E*}V_{kG}^{\bf i}}
{\epsilon_k-E_f}
\label{1.20}
\end{equation}
and 
\begin{equation}
{\cal B}_{\bf ii'}^{GE}=\bar{I}^{GE}_{\bf ii'}\Delta^G
\label{1.201}
\end{equation}
respectively. 
 
Then, one should diagonalize the quadratic form (\ref{1.11}), (\ref{1.19})
and obtain the spinon dispersion law $\varepsilon_{\beta}({\bf k})$ 
($\beta$ are the indices of spinon band) under the global constraint condition 
\begin{equation}
(2N)^{-1}\sum_{{\bf k},\beta}\left(
1-\tanh\frac{\varepsilon_{\beta}({\bf k})-\mu}{2T}
\right)= 1
\label{1.17}
\end{equation}
which is used in the mean-field approximation instead of exact local 
constraint (\ref{1.16}). This equation determines the spinon chemical 
potential $\mu$. 

Thus, to find the spin-fermion spectrum in the mean-field approximation 
we need only to diagonalize the Hamiltonian  (\ref{1.11}), (\ref{1.19})
under the constraint (\ref{1.17}). 
\section{Spinon spectrum of CeNiSn}

CeNiSn crystallizes in the orthorhombic lattice which 
belongs to the noncentrosymmetric space group $Pn2_1a$ \cite{Higash93}.
This structure can be described as zigzag chains of Ce atoms directed 
along $a$ axis (easy magnetization axis) and surrounded by slightly distorted 
trigonal "drums" formed by Ni and Sn ions. 
Thus, the point symmetry of the crystal field on Ce ions can be treated as 
nearly trigonal ($D_{3d}$) with rotation axis parallel to a-axis of the 
crystal, 
and the monoclinic distortion ($C_s$) can be considered as a small correction 
to the trigonal crystal field \cite{Aleks94}. Each elementary cell contains 
four Ce ions. 

To apply the theory of spin liquid state explicated in Section 2 to the case 
of CeNiSn one may use the irreducible representation of the trigonal point 
group $D_{3d}$ as a basis for Cornut-Coqblin model and then treat the $C_s$ 
distortion as a perturbation intermixing the trigonal CF terms. It was 
proposed in \cite{KKP93} (see also \cite{KVBT94}) and then confirmed 
in indirect experiment 
\cite{Aleks94} mentioned in the Introduction, 
that the ground state level and the first excited level 
form a pair of Kramers doublets 
\begin{eqnarray}
\left|G\pm \right\rangle &=&
a \left| \pm 1/2 \right\rangle \pm b \left| \mp 5/2 \right\rangle
\label{2.1a} \\
\left| E\pm \right\rangle &=& \left| \pm 3/2 \right\rangle
\label{2.1b}
\end{eqnarray}
separated by small energy interval which was estimated as $\approx 4.4$ meV 
in a point charge approximation for the crystal field.  Indeed, the purely 
electrostatic approximation is crude enough, because the covalent and exchange 
corrections should be taken into account as it is seen from our 
equations (\ref{1.12}) and (\ref{1.120}). These corrections result in 
additional reduction of the CF splitting (see below), 
so one can be sure that the third 
CF level is high enough in energy ($\approx 14.0$ meV according to 
calculations of \cite{Aleks94}), and the latter state is irrelevant 
to the low-energy excitation spectrum. 

To calculate the spectrum of RVB excitations one should diagonalize the 
mean field Hamiltonian $H_{MF}+\delta H_{MF}$ given by equations 
(\ref{1.11}) and (\ref{1.19}) by means of Fourier transformation in 
Ce-sublattice which, in turn, has four sublattices in this orthorhombic 
structure with the use of the basis functions (\ref{2.1a},\ref{2.1b}). 
This is a cumbersome procedure in  
general case, but in the special case of CeNiSn one should take into account 
extreme anisotropy of magnetic response which is seen both in static 
magnetic susceptibility \cite{Taka93} and in quasi 2D character of neutron 
scattering spectra \cite{Mason92,Kadow94}. 
These experimental facts enabled us to presume that the magnetic 
anisotropy reflects the quasi 2D character of low-energy spin excitations. 
and, hence, to consider the physical situation when these excitations  
possess the dispersion only in $bc$-plane. The 2D character of 
spin-fermion excitations can be explained by the properties of 
indirect RKKY interaction which is responsible for spinon coupling. 
It is obvious that only those 
spins can be involved in the resonating valence bond which are connected by 
the antiferromagnetic exchange coupling. If one adopts the positive
sign of the integral $I^{GG}_{{\bf i},{\bf i+\rho}}$ in $bc$ plane
($\rho$ is the distance between nn Ce ions), one can easily 
imagine that 15\% lesser value of intersite distance in $a$ direction 
\cite{Higash93} is 
enough to have nearly zero or even negative value of oscillating function  
$\Phi({\bf k}_F,{\bf R}_i-{\bf R}_{i+\rho})$ in the integral (\ref{1.10}). 
One should emphasize that this assumption does not imply  
two-dimensionality of conduction electron spectrum which can hardly be 
imagined in CeNiSn lattice. 

Another possible source of anisotropy of magnetic excitation spectrum is the 
anisotropy of indirect exchange described by the factor $B_G(\theta,\phi)$ in 
eq. (\ref{1.10}). The degree of anisotropy  given by this mechanism 
can be estimated  for a 
special case of the ground state (\ref{2.1a}) (see Appendix). 
This estimation shows that the "in plane" interaction exceeds 
the interattion along the $a$-axis   
when $|a| \gg |b|$ in the case of small interionic distances and when 
$|a| \ll |b|$ in the case of great interionic spaces. 
Since the interionic distance in CeNiSn is neither big nor
small and the experimental value of $a$ can be estimated as $a\approx 0.67 $ 
(see below), we have no real grounds for referring to this mechanism.

One more reason for changing the sign of $\Phi$ is the anisotropy of the 
electron Fermi surface. However, the available information about the latter 
is rather scanty, so we have no firm basis for discussing this
mechanism. 

In any case the assumption of two-dimensionality is not crucial for 
the results obtained, as soon as the exchange interaction along $a$-direction 
is not too big. The brief analysis of the influence of $a$-component of 
spinon dispersion on the neutron scattering spectra 
is presented in Subsection 5.4. 

Thus, assuming the 2D dispersion of spin liquid excitations,  we 
project the Ce-sublattice onto $bc$ plane and find that the 
orthorhombic 2D elementary cell contains two Ce ions in the sites 
${\bf i}={\bf l\xi}$ where $\xi=1,2$ defines the sublattice (see figure 1). 
This network is defined by the Bravais vectors ${\bf B} = (b,0)$ 
and ${\bf C} = (0,c)$ and the basis vector
${\bf d} = (0,-b/2,c/2-{\cal O})$. Here ${\cal O}$ is 
the orthorhombic distortion 
which transforms one-ion hexagonal lattice into two-ion orthorhombic one.

To describe the 2D spinon spectrum in a nearest neighbour approximation 
we define the coupling constants ${\cal T}_{\bf ii'}$ in  
equation (\ref{1.13}) by two parameters,
\begin{equation}
{\cal T}_{1}={\cal T}_{{\bf l}1,{\bf l'}1} ={\cal T}_{{\bf l}2,{\bf l'}2}
\label{2.201}
\end{equation}
for coupling within the same sublattice,  
and 
\begin{equation}
{\cal T}_2={\cal T}_{{\bf l}1,{\bf l'}2}={\cal T}_{{\bf l}2,{\bf l'}1}
\label{2.202}
\end{equation}
for intersublattice coupling. 
Similarly, two mixing integrals ${\cal B}_{\bf ii'}$ in equation 
(\ref{1.20}) are introduced, i.e. the one-cite coupling constant 
\footnote{This constant includes also the contribution of monoclinic 
distortion of electrostatic crystal field} 
\begin{equation}
{\cal G}_1={\cal B}^{GE}_{{\bf l}1,{\bf l}1}=  
{\cal B}^{GE}_{{\bf l}2,{\bf l}2}\;.
\label{2.301}
\end{equation} 
and inter-site coupling constant 
\begin{equation}
{\cal G}_2={\cal B}^{GE}_{{\bf l}1,{\bf l'}2}= 
{\cal B}^{GE}_{{\bf l}2,{\bf l'}1}\;.
\label{2.302}
\end{equation}  
Such choice implies that the very existence of two sublattices is due to 
orthorhombic distortion ${\cal O}$ of trigonal lattice, so 
displacement of second sublattice (see figure 1) is, probably, the 
main source of the non-Coqblin hybridization. 
Then, introducing the Fourier transformation of spinon operator, 
\begin{equation}
f_{{\bf k}\mu} = N^{-1/2}\sum_{\bf l}
\sum_{\xi = 1}^{2} \sum_{\Gamma}^{G,E} \sum_{\nu}^{\pm}   
\Xi_{\mu}^{\Gamma,\nu} (\xi,{\bf k})
f_{{\bf l}\xi,\Gamma\nu}\exp{(i{\bf kl})}
\label{2.4}
\end{equation}
and taking into account the fact that this type of displacement generates 
the crystal field which components are given by the Stevens operators 
$\hat{O}_n^2$ (see also Subsection 5.2), 
we find the system of equations  for the coefficients 
$\Xi_{\mu}^{\Gamma,\nu} (\xi,{\bf k})$ and eigenvalues
$\varepsilon_{\mu} ({\bf k})$ , 
\begin{equation}
\sum_{{\bf l}'}^{N} \sum_{\xi' = 1}^{2} 
\sum_{\Gamma'}^{G,E} \sum_{\nu'}^{\pm}   
{\cal D}_{\Gamma\Gamma';\nu\nu'}^{\xi\xi'}({\bf R}_{ll'})
\Xi_{\mu}^{\Gamma',\nu'} (\xi',{\bf k})
\exp \left\{ i\bf{kl}' \right\}
=
\varepsilon_{\mu} ({\bf k}) 
\Xi_{\mu}^{\Gamma,\nu} (\xi,{\bf k})
\exp \left\{ i\bf{kl} \right\}\;.
\label{2.5}
\end{equation}
Here the coefficients 
$D_{\Gamma\Gamma',\nu\nu'}^{\xi\xi'}({\bf R}_{ll'})$ are defined as 
\begin{eqnarray}
{\cal D}_{EE;\nu\nu}^{\xi\xi}({\bf R}=0) =\widetilde{\Delta}_{CF}, \;\;\; 
{\cal D}_{GG;\nu\nu}^{11}({\bf R}={\bf R}_1)= {\cal T}_{1},\;\;\;
{\cal D}_{GG;\nu\nu}^{12}({\bf R}={\bf R}_1)= {\cal T}_{2},\;\;\;
\nonumber \\
{\cal D}_{EG;\nu\nu'}^{11}({\bf R}=0) = 
{\cal G}_1 \;\; \;\
{\cal D}_{EG;\nu\nu'}^{12}({\bf R}={\bf R}_1) = 
{\cal G}_2
\label{2.6}
\end{eqnarray}
\sloppypar
[see equations (\ref{2.201}) -- (\ref{2.302})], 
${\bf R}_1$ is the coordinate of the nearest neighboring site.   
Then the secular matrix can be decoupled into two blocks,   
$\left\{(1G+), (2G+), (1E-), (2E-) \right\}$ and 
$\left\{(1G-), (2G-), (1E+), (2E+) \right\}$
which correspond to two new Kramers doublets given by the following 
linear combinations, 
\begin{equation}
f_{{\bf k}\mu,\pm,} =
\Xi_{\mu}^{G} (1,{\bf k}) f_{{\bf k}1,G\pm} +
\Xi_{\mu}^{G} (2,{\bf k}) f_{{\bf k}2,G\pm} +
\Xi_{\mu}^{E} (1,{\bf k}) f_{{\bf k}1,E\mp} +
\Xi_{\mu}^{E} (2,{\bf k}) f_{{\bf k}2,\mp} \;.
\label{2.7}
\end{equation}
Each block of the secular matrix has the form 
\begin{eqnarray}
\left(
\begin{array}{cccc} 
 {\cal T}_{1}M_1({\bf k})  & {\cal T}_{2}M_2({\bf k}) & {\cal G}_1 & {\cal G}_{2}M_2({\bf k}) \\ 
 {\cal T}_{2}M^*_2({\bf k}) & {\cal T}_{1}M_1({\bf k}) & {\cal G}_{2}M^*_2({\bf k}) & {\cal G}_1  \\ 
 {\cal G}_1 & {\cal G}_{2}M_2({\bf k}) & \widetilde{\Delta}_{CF} & 0      \\   
 {\cal G}_{2}M^*_2({\bf k}) & {\cal G}_1 & 0      & \widetilde{\Delta}_{CF}  
\end{array}
\right)
\label{2.8}
\end{eqnarray}
The structure factors which define the spinon dispersion are determined as    
\begin{equation}
M_1({\bf k}) = 2\cos \left( {\bf kB} \right)
\label{2.9}
\end{equation}
\begin{equation}
M_2({\bf k}) = 4 \exp \left( i \frac{{\bf kB} - {\bf kC}}{2} \right)
\cos \left( \frac{{\bf kB}}{2} \right)
\cos \left( \frac{{\bf kC}}{2} \right)
\label{2.10}
\end{equation}
(${\bf B} = (b,0)$ and ${\bf C} = (0,c)$). 

The excitation spectrum  is shown in figure 2 (the insert  
presents the dispersion of spin excitations in 
Coqblin-Cornut approximation $({\cal G}_1=0, {\cal G}_{2}=0)$ 
where the CF excitations 
are simply superimposed on the ground state spinon band). 
It is seen from this figure that the intermixing of spinon and CF excitations 
results in radical reconstruction of the low-energy part of this spectrum. 
We associate the characteristic temperature $T^*$  estimated from various 
experiments as $T^*\approx 80$K  (see Introduction) with the 
characteristic energy scale of this structured part, so   
the values of ${\cal T}_1 = 12.7$K and ${\cal T}_2 = 18$K were chosen 
to reproduce approximately the width of this energy interval. The values  
of other parameters (${\cal G}_1$, ${\cal G}_2$ and $\widetilde{\Delta}_{CF}$) 
where taken to fit the positions of the main peaks in the neutron 
scattering  spectra (see the next section). In particular, we adopted the 
value of 13.5 K for the renormalized CF splitting parameter 
$\widetilde{\Delta}_{CF}$). Its reduction in comparison with the value of 
48.4 K given by indirect experiment of \cite{Aleks94} is easily explained by 
the effects of covalent and exchange renormalization given by the terms  
$B_c^{G}$, $B_c^{E}$ and $B_{ex}^{G}$ in eqs. 
(\ref{1.12}) and (\ref{1.120}). 
It is obvious that the covalent repulsion 
$B_c^{\Gamma}$ of the localized levels $E_{\Gamma}$ from the  
free conduction band states above the Fermi level (\ref{1.12},\ref{1.120})
is governed by the value of hybridization matrix elements 
$|V_{k\Gamma}^{\bf i}|^2$. 
Since the Fermi surface of CeNiSn is dominated by the 
$p$-partial waves \cite{Noha93}, the hybridization is largest for 
the $j_z=\pm 1/2$ and $j_z=\pm 3/2$ components of the f-states. 
Therefore,  
the ratio of hybridization matrix elements for $\left|G\pm \right\rangle$ 
(\ref{2.1a}) and $\left|E\pm \right\rangle$ (\ref{2.1b}) can be evaluated as 
$|V_{kG\pm}^{\bf i}|^2 / |V_{kE\pm}^{\bf i}|^2 \approx a^2 < 1$. 
This means that the {\it negative} covalent term reduces the  
CF splitting. 
The exchange contribution $B_{ex}^G$ in 
(\ref{1.12}) gives additional contribution to this reduction effect. 
\par
The dispersionless behavior of spinon spectrum at the $b$-facet
of Brillouin zone $[(0,1/2,0)\leq {\bf k} \leq (0,1/2,1/2)$ ] is explained 
by the specific form of the spinon spectrum 
\begin{equation}
\varepsilon_{\mu= 1,2} ({\bf k}) =  
2\left\{{\cal T}_{1}
\cos \left( {\bf kB} \right) \pm 
2{\cal T}_{2}
\cos \left( \frac{{\bf kB}}{2} \right) 
\cos \left( \frac{{\bf kC}}{2} \right)
\right\}
\label{2.11}
\end{equation}
since this spectrum becomes dispersionless for ${\bf kB} = \pi$. 

It is important that inclusion of intermixing terms ${\cal G}_{1,2}$ in the 
secular matrix (\ref{2.8}) does not alter the one-dimensionality 
of this branch (cf. upper and lower panels of figure 2) 
because the form factor $M_2$ given by eq. (\ref{2.10}) also turns into
zero at ${\bf kB} = \pi$. Basing on the spinon dispersion law shown 
in this picture we will try in the next section to explain the neutron 
scattering data. 
The chemical potential of spin-fermions at $T=0$
which is determined by the constraint (\ref{1.17}) is shown in this picture 
by the dashed line. 

Thus we find that the spin excitations in CeNiSn preserve to some extent 
the features of the one-site crystal field excitation spectrum: the multiband 
dispersion picture contains vast dispersionless parts which origin is the 
initial atomic crystal field splitting of the f-level. The density of 
spin-fermion states described by these bands is shown in upper and 
lower panel of figure 3. 
We see that instead of simple two-level picture of CF excitations there 
exists several peaks some of which can be considered as the  remnants of 
CF levels but other reflect the van Hove singularities of 2D spinon spectrum. 
The spectrum demonstrates the pseudogap features presupposed in 
early phenomenological descriptions \cite{Taka93}. However, it is   
remarkable that the chemical potential of spin-fermion 
excitations falls not into the 
hybridization gap but into the pseudogap between two van Hove peaks. 
\section{Inelastic Neutron Scattering in Spin Liquid}
The inelastic neutron scattering provides unique experimental information 
concerning the structure and dispersion of low-energy spin excitations, 
so these data look challenging for the general theory of spin liquid 
and especially for the case of interplay between HF anf CF excitations.
To calculate the spectra of inelastic neutron scattering in the spin 
liquid,  we start with the well 
known equation for magnetic scattering cross section 
\begin{equation}
\frac{d^2 \sigma}{d \Omega d E'} = 
\left( \frac{\gamma e^2}{m_e c^2}  \right)^2 \frac{\kappa'}{\kappa}
\sum_{\alpha \beta} 
\left( \delta_{\alpha \beta} - \hat{Q}_{\alpha} \hat{Q}_{\beta} \right)
S_{\alpha \beta} ({\bf Q}, \hbar \omega)
\label{3.1}
\end{equation}
Here $m_e$ and $e$ are the mass and the charge of electron, $\gamma = -1.91$
is the gyromagnetic ratio for neutron, $c$ is the light velocity.
The wave vectors $\kappa$ and $\kappa'$ are momenta of incident and 
scattered neutron,  respectively,  and 
$\hat{Q}_{\alpha}$ are the cartesian components of unit vector directed 
along the momentum transfer vector ${\bf Q}$. The component 
$S_{\alpha \beta} ({\bf Q}, \hbar \omega)$ of the scatering function can be 
expressed as the sum over all possible initial ($\lambda$) and final 
($\lambda'$) states with the energies $E_{\lambda}$ and $E_{\lambda}$, 
respectively, 
\begin{equation}
S_{\alpha \beta} ({\bf Q}, \hbar \omega) =
\sum_{\lambda\lambda'} n_{\lambda} 
\left\langle \lambda \left| \hat{\cal Q}_{\alpha}^+ \right| \lambda' \right\rangle 
\left\langle \lambda' \left| \hat{\cal Q}_{\beta} \right| \lambda \right\rangle 
\delta \left( \hbar \omega + E_{\lambda} - E_{\lambda'} \right)
\label{3.2}
\end{equation}
($n_{\lambda}$ is the probability distribution for initial state). 
Then, the operator $\hat{\cal Q}$ of neutron-electron interaction
can be expressed as a sum over all f ions, 
\begin{equation}
\hat{{\cal Q}} = \sum_{\bf l} \exp \left (i {\bf Q} {\bf R}_l \right) 
\sum_{\xi}^{\mbox{cell}} \exp \left (i {\bf Q} {\bf d}_{\xi} \right) 
\hat{{\cal Q}}_{l \xi}
\label{3.3}
\end{equation}

In the dipolar approximation the operator $\hat{{\cal Q}}_{l \xi}$ turns into 
\begin{equation}
\hat{{\cal Q}}_{l \xi} = \frac{1}{2} g F({\bf Q}) \hat{{\bf J}}_{l \xi}
\label{3.30}
\end{equation}
where g is the Lande splitting factor, $F({\bf Q})$ is the ionic fomfactor and 
$\hat{{\bf J}}$ is the total angular momentum operator. 
Thus the problem is reduced to calculation of the correlator 
$\left\langle \lambda \left| \hat{J}_{l \xi}^+  
\hat{J}_{l' \xi'}^- \right| \lambda \right\rangle $

In our model the source of neutron scattering are the transitions 
between the spin liquid excitations described by the operators (\ref{2.4})
with the dispersion law $\varepsilon_{\mu}({\bf k})$ 
given by the secular equation  (\ref{2.8}). 
Now the situation formally resembles the well-known paramagnetic 
scattering \cite{Doni67}
where the spin-fermion excitations $\lambda={\bf k}\mu$ play the same part as 
the itinerant electrons in transition metals.   
Substituting the eigen vectors given by equation (\ref{2.4}) in
the scattering function (\ref{3.2}), we find
\begin{eqnarray}
&S_{\alpha \beta} ({\bf Q}, \hbar \omega) = 
\left( \frac{1}{2} g F({\bf Q}) \right)^2 &\nonumber \\
&\sum_{\mu\mu'} \sum_{\bf k} 
\left[ I_{\alpha}^{\mu\mu'} ({\bf k}, {\bf Q}) \right]^*
I_{\beta}^{\mu\mu'} ({\bf k}, {\bf Q}) 
n_{\mu}({\bf k}) \left[ 1 - n_{\mu'}({\bf k}-{\bf Q}) \right]
\delta \left( \hbar \omega + \varepsilon_{\mu} ({\bf k}) - 
\varepsilon_{\mu} ({\bf k}-{\bf Q}) \right)&
\label{3.4}
\end{eqnarray}
where 
\begin{equation}
I_{\alpha}^{\mu\mu'} ({\bf k}, {\bf Q}) = 
\sum_{\Gamma\nu} \sum_{\Gamma'\nu'} \sum_{\xi}
\exp \left (i {\bf Q} {\bf d}_{\xi} \right) 
\left( \Xi_{\mu}^{\Gamma,\nu} (\xi,{\bf k}) \right)^*
\Xi_{\mu'}^{\Gamma',\nu'} (\xi,{\bf k}- {\bf Q})
\left\langle \Gamma \nu \left| \hat{J}^{\alpha}  \right| \Gamma' \nu' \right\rangle
\label{3.5}
\end{equation}
$n_{\mu}({\bf k})$ is the Fermi distribution function for spinon excitations
\par 
This expression can be simplified in case of CeNiSn. It is 
known from experiment \cite{Sato95,Kambe96} that the dominant component which 
contributes significantly to the scattering function is 
$S_{aa} ({\bf Q}, \hbar \omega)$ 
component. Thus, for the momentum transfer ${\bf Q}$ in the $b-c$ plain,
which is the experimental condition for the most of experimental scans, 
one has for the scattering function 
$$
S_{aa} ({\bf Q}, \hbar \omega) =
\left( \frac{1}{2} g F({\bf Q}) \right)^2 
$$
\begin{equation}
\sum_{\mu\mu'} \sum_{\bf k} 
\left| I_{a}^{\mu\mu'} ({\bf k}, {\bf Q}) \right|^2
n_{\mu}({\bf k}) \left[ 1 - n_{\mu'}({\bf k}-{\bf Q}) \right]
\delta \left( \hbar \omega + \varepsilon_{\mu} ({\bf k}) - 
\varepsilon_{\mu} ({\bf k}-{\bf Q}) \right)
\label{3.6}
\end{equation}
Since the axis $a$ is the axis of spin quantatisation, 
and since the operator $J_a$ has only diagonal 
nonzero matrix elements in the basis of CF states,   
$\left| \Gamma \nu \right\rangle$ (\ref{2.1a},\ref{2.1b}), the expression for  
$I_{\alpha}^{\mu\mu'} ({\bf k}, {\bf Q})$ has the form 
\begin{equation}
I_{a}^{\mu\mu'} ({\bf k}, {\bf Q}) = 
\sum_{\Gamma\nu} \sum_{\xi}
\exp \left (i {\bf Q} {\bf d}_{\xi} \right) 
\left( \Xi_{\mu}^{\Gamma,\nu} (\xi,{\bf k}) \right)^*
\Xi_{\mu'}^{\Gamma,\nu} (\xi,{\bf k}- {\bf Q})
\left\langle \Gamma \nu \left| \hat{J}^{a}  \right| \Gamma \nu \right\rangle
\label{3.7}
\end{equation}
Finally, taking into account the explicit equation (\ref{2.7}) for the  
spinon operator and expressions for the CF states  
(\ref{2.1a},\ref{2.1b}) one comes to the final equation for the matrix element 
$I_{\alpha\pm}^{\mu\mu'} ({\bf k}, {\bf Q})$ (signs $\pm$ are for up and down 
partner of the Kramers doublet (\ref{2.7}) respectively)
\begin{eqnarray}
I_{\alpha\pm}^{\mu\mu'} ({\bf k}, {\bf Q})& = & 
\left[
\left( \Xi_{\mu}^{G} (1,{\bf k}) \right)^* 
\Xi_{\mu'}^{G} (1,{\bf k}- {\bf Q}) \Theta_{G\pm}
+
\left( \Xi_{\mu}^{E} (1,{\bf k}) \right)^* 
\Xi_{\mu'}^{E} (1,{\bf k}- {\bf Q}) \Theta_{E\mp}
\right] +
\nonumber \\
&& 
\left[
\left( \Xi_{\mu}^{G} (2,{\bf k}) \right)^* 
\Xi_{\mu'}^{G} (2,{\bf k}- {\bf Q}) \Theta_{G\pm}
+
\left( \Xi_{\mu}^{E} (2,{\bf k}) \right)^* 
\Xi_{\mu'}^{E} (2,{\bf k}- {\bf Q}) \Theta_{E\mp}
\right]  \nonumber \\
&&\times \exp \left( i {\bf Q} {\bf d}  \right)
\label{3.8}
\end{eqnarray}
where 
\begin{equation}
\Theta_{G\pm} = \pm \left(3a^2- \frac{5}{2}\right)
\end{equation}
\begin{equation}
\Theta_{E\pm} = \pm \frac{3}{2}
\label{3.9}
\end{equation}
\section{Inelastic neutron scattering spectra in CeNiSn}
In this section  we compare the inelastic neutron  scattering spectra
calculated by means of eqs. (\ref{3.6}) and (\ref{3.8}) with 
dispersion law for spin-fermions given by secular matrix (\ref{2.8}) 
and figure 2 with the experimental data of the refs. 
\cite{Mason92,Kadow94,Sato95,Kambe96}  

\subsection{Qualitative considerations}

The inelastic neutron scattering spectrum of CeNiSn in its gross features 
is formed by two main signals with the energies of $\sim 2$meV and $\sim4$meV. 
It is seen from the lower panels of figures 2 and 3 that these "peaks" 
can be ascribed to transitions from the occupied states in the lower 
band to the region of the empty levels where the states $|G\pm\rangle$ 
(\ref{2.1a}) and $|E\pm\rangle$ (\ref{2.1b}) are strongly hybridized. 
The most distinctly these features are seen in experiment 
for the momentum transfer vector 
${\bf Q}$ in the $bc$-plain and, hence, the magnetic response 
is connected with the $\chi^{aa}$ component of dynamic magnetic susceptibility
\cite{Sato95}.  
As is known from the measurements of static succeptibility,  
the $a$-axis is the easy magnetization axis in CeNiSn \cite{Taka93}. 
Since the lower spinon band is formed mainly by the ground state doublet 
$|G\pm\rangle$ (\ref{2.1a}) where the components $|\pm 5/2\rangle$ 
play the main part (see next subsection), the longitudinal 
susceptibility should be the strongest component of  magnetic response at 
low energies, so the dominance of 
${\bf Q}\perp a$ momentum transfer vectors is naturally explained in 
our model. Then the most intense transitions ar expected for those final 
states where the greater fraction of $|\pm 5/2\rangle$ is admixed to the 
$|\pm 3/2\rangle$ states. 
 
\par
The 2meV peaks 
are seen experimentally for ${\bf Q}=(0,0,1)$ \cite{Mason92} and 
${\bf Q}=(0,1,0)$ \cite{Kadow94,Sato95}. 
The corresponding momentum transfer vectors are indicated by solid arrows in 
figure 4 where the shaded area is the projection of Brillouin 
zone to the $bc$ plane).
These processes correspond to the vertical transitions from the spinon 
"Fermi surface" to the region of strong hybridization near the point 
${\bf k}=(0,0,1/2)$. The bold 
arrow marks these transitions in figure 2  
(the integer reciprocal vector (0,1,0) or (0,0,1) are substracted). 
The same transitions are shown by the bold arrow in the lower panel 
of figure 3. The highest peak in DOS  
arises due to the van Hove singularities near the former CF excitation 
level. 
\par
The peaks at $E=4$meV are seen for ${\bf Q}=(0,1/2+n,Q_c)$ 
(grey arrows in figure 4) where $n$ are integer numbers \cite{Kadow94,Sato95}. 
This maximum in the scattering cross section is due to the intraband 
transitions shown by grey arrows in figure 2.
It is easily seen that since the 4meV peak corresponds to the transitions 
between two peaks of DOS (grey arrow in the lower panel of figure 3), 
and the 2meV processes are due to the transitions 
from the structurless part of DOS to the empty peak, the intensity of the 
former signal should be higher than that of the latter one.
This result is also consistent with the experimental data 
\cite{Kadow94,Sato95}. 
\par
It worth to be noted that the general shape of the DOS in the lower band 
reminds the double-peak DOS with pseudogap  postulated 
in early phenomenological models \cite{Taka93,KKP93,Nakam95}. However, 
at least {\it one more} peak with higher energy should exist in the 
scattering spectrum given by the present model. This peak is formed by the  
interband transitions shown by the dashed lines in figures 2 and 3. 
Such peak with $E\approx 7$meV was found in the recent neutron 
scattering experiments \cite{Kambe96}. 
\par
Thus, the present model explains all qualitative features  of the 
neutron scattering spectrum. To make the detailed quantitative 
description of experimental scans one should first specify the model 
parameters. 

\subsection{Evaluation of the model parameters}

To calculate the spinon spectrum determined by the secular matrix (\ref{2.8})
one should know the values of 
the intersite coupling constant ${\cal T}_{1,2}$, 
the mixing parameters ${\cal G}_{1,2}$ the energy 
$\widetilde{\Delta}_{CF}$ of CF splitting. 
To find the intensities of neutron scattering 
one needs also the magnitude of the orthorhombic distortion ${\cal O}$
and the values of parameters $a,b$ which determine the structure of the 
ground state CF level $|G\pm\rangle$ (\ref{2.1a}). All these quantities, 
except hybridization parameters ${\cal G}_{bc}$ and ${\cal G}_{0}$ are, indeed,
predetermined by the experimental data available.
The criterions for choosing the parameters ${\cal T}_{1,2}$ and 
$\widetilde{\Delta}_{CF}$ are described in Section 3. 
The value of orthorhombic distortion ${\cal O}$ in units of lattice parameter 
$c$ is 0.10 \cite{Higash93}.
\par
We have no independent experimental information to choose the values of  
mixing parameters ${\cal G}_{1,2}$, so we use these 
parameters for fitting the experimental neutron scattering cross section. 
The best values, which gave us the possibility to describe the low-energy 
magnetic scattering function for all measured momentum transfer 
are ${\cal G}_{1} =-3.5$ K, ${\cal G}_{2} =-7.7$ K. 
\par
To estimate the parameters $a,b$ entering the ground state CF level wave 
function (\ref{2.1a}) we used the experimental data on the anizotropy of 
low temperature static magnetic susceptibility at low temperatures
\cite{Taka96}. According to our model, this susceptibility should be 
mainly determined by the Pauli-like contribution of RVB excitations. 
\par
The Pauli-like contribution for magnetic field applied along the $j$-axis 
at low 
temperatures is proportional to the DOS of spinons $N(\varepsilon_F)$ 
and to the square of effective $g^{(p)}_j$-factor of the spinons states
$|\widetilde{G}\pm\rangle$ at the Fermi level, 
$\chi^{(p)}_{x} \sim  \left| g_x^{(p)}  \right|^2 {\cal N}(\varepsilon_F)$ 
(${\cal N}(\varepsilon_F)$ is the density of spinon states at the Fermi level).
The effective $g^{(p)}_j$-factor is determined as 
$g^{(p)}_j = 2 g_{J} 
| \langle \widetilde{G} \nu | \hat{J}_j | \widetilde{G} \nu' \rangle |$
($\nu=\pm$). Here $g_{J}=6/7$ is the Lande factor for Ce$^{3+}$ and
$\hat{J}_j$ is the $j$- component of total momentum operator. 
According to (\ref{2.7}), 
the spinon states at the Fermi surface  
which determine the anisotropy of the susceptibility are described as 
the mixture of the bare wave functions 
$|G\pm\rangle$ and $|E\pm\rangle$ (\ref{2.1a}),(\ref{2.1b}). 
due to orthorhombic
distortion of the trigonal lattice. 
As was mentioned above, we consider the $\hat{O}_n^2$ 
orthorhombic distortion (see Fig. 1). 
Therefore, the wavefunctions in the vicinity of the Fermi surface can be 
approximated by 
\begin{equation}
|\widetilde{G}\pm\rangle = 
\alpha_{av}^{(F)} |G\pm\rangle + \beta_{av}^{(F)} |E\mp\rangle
\label{6.1}
\end{equation}
where $\alpha_{av}^{(F)}$ and $\beta_{av}^{(F)}$ 
are the mixing parameters averaged 
over the Fermi surface.
\par
When evaluating the Pauli contribution, we  
neglect the anisotropy of spinon Fermi surface 
and assume that the ratio between the components of static susceptibility 
is given by the anisotropy of $g$ factor:
$$\chi^{a}\;:\;\chi^{b}\;:\;\chi^{c} \approx 
|g_a|^2\;:\;|g_b|^2\;:\;|g_c|^2.$$
To evaluate the averaged mixing parameters 
$\alpha_{av}^{(F)}$ and $\beta_{av}^{(F)}$ we take into account 
that there are $N=4$ nearest neighbors from the second sublattice which take 
part in k-dependent hybridization. The averaging of the k-dependence of 
hybridization gives additional factor $1/2$. The distance from the Fermi 
surface $\varepsilon_{F}$ to the crystal field level is 
$\approx 30$ K according to our calculations (see figures 2,3). 
Therefore, the value of $\beta_{av}$ at the Fermi 
surface can be estimated as 
$\beta_{av}^{(F)}  \approx N{\cal G}/2(\Delta_{CF} - \varepsilon_{F})$ ,
so  $\alpha_{av} \approx 0.88$.
The similar value of parameter parameter $\alpha_{av}$ can be extracted from 
the experimental data for the anisotropy of magnetic susceptibility
\begin{equation}
s_1 = \frac{\left|\chi_b - \chi_c \right|}{\chi_c} \approx 
\left( \frac{4\sqrt{2}\sqrt{1-\alpha_{av}^2}}{3\alpha_{av}a} \right)^2
\label{6.3}
\end{equation}
(where $s_1 \approx 0.25$ \cite{Taka96}). One finds 
$\alpha_{av} \approx 0.94$ from (\ref{6.3}) in agreement with the 
theoretical evaluation.
The estimation of the magnetic susceptibility gives 
\begin{equation}
s_0 \equiv \frac{\chi_a}{\chi_c} \approx \frac{\chi_a}{\chi_b} \approx 
\left( \frac{\alpha_{av} \left(2-6a^2 \right)+3}{3\alpha_{av}^2 a^2} \right)^2
\label{6.2}
\end{equation}
where $2 < s_0 <3$.  Using (\ref{6.2}) the value of $|a|$ is determined
as 
\begin{equation}
|a| \approx \sqrt{\frac{2\alpha_{av}^2+3}{3\alpha_{av}^2(2+\sqrt{s_0})}}
\end{equation}
so we obtain  $0.65< |a| < 0.75$.
Since the contribution of $|5/2\rangle$-component in the magnetic response
is dominant even for $|a| \approx |b|$, 
we actually deal with the $|\pm5/2\rangle$ states 
in the  neutron scattering experiments. 
It turns out that the neutron scattering cross sections weakly depend on the 
the value of $a$ in the actual range of $|a|$, and we can not extract this 
parameter from the available experimental data. 
In the following calculations the value of $|a| = 0.67$ is taken.  

\subsection{Quantitative description  of constant-${\bf Q}$ and constant-E 
neutron scattering scans} 

First, we see that our model reproduces quite well the "pseudogap" behavior of 
constant-${\bf Q}$ scans first found in \cite{Mason92}. Figure 5 presents 
the theoretical curves for ${\bf Q}$ perpendicular to ${\bf a}$ axis in 
comparison with experimental data of ref. \cite{Mason92} and later results 
of \cite{Kadow94,Sato95}. It is easily seen from Figs 2 and 3 (lower panel)  
that this character of ${\bf Q}\perp{\bf a}$ scans is due to high density of
states for the $|\pm 5/2\rangle$ 
components in the region of strong hybridization around $(0,0,1/2)$ point. 
This psedogap-like behavior disappears for  ${\bf Q}\|{\bf a}$ \cite{Mason92}  
simply because the 2-meV peak 
is suppressed for this direction of momentum transfer in accordance with 
the geometrical factor in eq. (\ref{3.1}) under condition
$\chi_{aa}\gg \chi_{bb},\chi_{cc}$. 

The next striking feature of experimental scans is the quasi 1D character of 
4-meV excitation: it was noticed \cite{Kadow94,Sato95} that the behavior 
of this peak for ${\bf Q} = (0,n+1/2,Q_c)$ practically does not depend 
on the value of $Q_c$. Our next scan demonstrates, first, that this peak 
is indeed can be obtained in the spin-fermion model (figure 6, upper 
panel) and, second, that nothing is changed significantly 
in the shape of this scan with varying $Q_c$
(lower panel). 
The form of spinon dispersion curve explains this one-dimensionality:
due to the dispersionless character of the energy band along the wave 
vector ${\bf k}$ in ${\bf k}= (0,1/2,k_c)$ direction 
[see equation (\ref{2.11}) for explanation], 
the contribution from the intraband transitions with $E\sim 4$meV 
to the scattering function (spanted arrows in fiture 2b) does not
depend on $Q_c$ and thus mimics quasi one-dimensional scattering.       

One more remarkable property of experimental constant-${\bf Q}$ scans 
is their modification along the $(0,Q_b,0)$ direction. It was found 
(see \cite{Sato95}, figure 3) that the form of 
these scans change from the structure with definite 4.2 meV peak at 
$Q_b=1.5$ to that with 2 Mev peak at $Q_b=1$. It is seen from figure 7 that 
our theoretical scans reproduce this variation quite  satisfactorily, 
including the featureless shape of the scan at intermediate value of 
$Q_b=1.2$.

The theory explains also why the 2 meV peak  which is distinctly seen for 
the vertical transitions $(0,Q_b,0)$ and $(0,0,Q_b)$ (see figures 4,5) 
practically disappears for the "diagonal" direction ${\bf Q}=(0,1,1)$ 
(figure 16 of \cite{Sato95}). This kind of anisotropy is explained by the
fact that  the phase factor 
$ \exp \left( i {\bf Q} {\bf d}  \right)$
in (\ref{3.8}) is close to unity for ${\bf Q} = (0,1,1)$
due to the small orthorhombic shift ${\cal O} = 0.10$ (see Subsection 5.2).
Then, since the condition $\Theta_{G\pm}\approx \Theta_{E\mp}$ is valid 
for the chose value of $a=0.67$,
the matrix element (\ref{3.7}), (\ref{3.8}) 
practically turns out into the orthogonality relation 
between the different quantum states (\ref{2.4}) with the same reduced wave 
vector. This is why the intensity of scan at ${\bf Q} = (0,1,1)$
is by order of magnitude smaller than that of ${\bf Q} = (0,0,1)$ or
${\bf Q} = (0,1,0)$. 

Finally, we discuss the characteristic features of constant-E scans. 
The calculated scans for $E=4.2$meV are presented in figure 8 together 
with the experimental data taken from \cite{Sato95}. It is 
no wonder that the theoretical scans are in satisfactory 
quantitative agreement with experiment: it simply follows from above
explanation that the maxima of $(0,Q_b,0)$ and $(0,Q_b,1)$ scans  
at $Q_b = 0.5,1.5$ are connected with the fact 
that the 4meV peak arises just at ${\bf Q} = (0,n+1/2,x)$ 
with integer $n$ and any $x$.

More puzzling feature of the constant-E scans was noticed in \cite{Kambe96} 
where it was observed that the peak at
$Q_b=1.5$ for the $(0,Q_b,1)$ scans with $E=4.2$meV splits into two peaks for 
lower energy transfer $E=3.3$meV and $E=2.5$meV. 
This feature is reproduced, at least quantitatively, in our 
calculations (see figure 9). Both the appearance of the doubled peak and the 
increase of the splitting for lower energies is seen in the 
theoretical curves. 
Even the higher intensity of the left peak is reproduced in the 
calculated scan. 
To investigate the reason of this asymmetry we calculated also the 
constant-{\bf Q} scans with ${\bf Q}=(0,Q_b,1)$ for $Q_b = 1.25, \; 1.5 \; 1.75$. 
Our figure 10 demonstrates that for higher energy transfer the intensity of
$Q_b = 1.5$ response is higher than that of $Q_b = 1.25, \; 1.75$ whereas
for lower energy transfer the situation is reversed. 

Thus, the totality of calculated scans shows that not only the general 
shape but also the anisotropy of spinon spectrum correlates with the 
details of experimental neutron scattering scans. These results 
demonstrate clearly 
that the anisotropic magnetic response of CeNiSn is due to anisotropy of 
the spectrum of low-energy spin-fermion excitations in this system.
\subsection{Influence of 3-dimensionality of the spinon spectrum}

To check the robustness of our results to assumption about mainly 
two-dimensional character of spinon dispersion we recalculated 
the spinon DOS and inelastic neutron scattering intensities. 

We introduced in the secular matrix for the spinon spectrum two more 
parameters, i.e. the "overlap integral" 
${\cal T}_3={\cal T}_{{\bf l}1,{\bf l'}3}\equiv{\cal T}_{{\bf l}2,{\bf l'}4}$ 
where the indices (3,4) stand for the Ce ions above and below the sites 
(1,2) respectively, in $a$-direction, and 
${\cal G}_3={\cal B}_{{\bf l}1,{\bf l'}3}^{GE}\equiv
{\cal B}_{{\bf l}2,{\bf l'}4}^{GE}$ with the same notations. 
These parameters result in further doubling of the crystal cell 
lattice and further removing  the degeneracy of spinon spectrum and 
appearance of spinon dispersion in $a$-direction of the Brillouin zone. 
As to the spinon DOS, the overlap parameter ${\cal T}_3$ is responsible mainly 
for for broadening of the second van Hove peak (-25 K peak in the lower 
panel of fig.3) and the influence of the hybridization parameter ${\cal G}_3$  
is seen in broadening and  lowering of the 3 K peak in the mentioned DOS. 
But the general shape of the DOS, and, therefore, the form of the 
inelastic neutron scattering spectra for ${\bf Q}$ in $bc$-plane 
are practically not modified at least for ${\cal T}_3$ and ${\cal G}_3$ less
then $\approx 4K$, i.e. more then 50 per cent of the characteristic values 
of in-plane interaction parameters. 
Correspondingly, the neutron scattering peaks for the momentum transfer in 
$bc$ plane  are also slightly broadened, but the general quantitative agreement 
is still satisfactory, although the CF splitting parameter should be 
slightly changed. If one chooses, e.g., ${\cal T}_3 = {\cal G}_3 = 3.5 $K, the 
renormalized value of $\tilde{\Delta}_{CF}=19.5$K should be taken to 
get an agreement with the experimental data for all scans demonstrated in 
figures 5-10. 

However, the introduction of third component in spinon spectrum provided us 
one more possibility to check the model assumptions, i.e. to calculate 
the neutron scattering for ${\bf Q}$ out of $bc$-plane. In fig.11 the 
constant-E scan for ${\bf Q}=(Q_a, 3/2, 0)$ is presented in comparison with 
the experimental data of \cite{Sato95}. The oscillating character of 
scattering intensity with $Q_a$ is distinctly seen in experiment and is 
shurely reproduced in the calculated curve. It is interesting that the 
shape of the theoretical curve is more complicated than the rough 
sinusoidal guideline shown in fig.11d of Ref. \cite{Sato95}, and the 
attentive look on the experimental points confirms this picture at 
least qualitatively. 

When calculating the theoretical curve we weighted the inelastic 
cross section only with the form factor $F({\bf Q})$ 
[see eq. (\ref{3.4})]. This procedure resulted in better agreement 
with the experiment than that including also the orientational factor 
$1-\hat{Q}_a^2$. This means that the contribution of ${\rm Im}\chi^{bb}$ and 
${\rm Im}\chi^{cc}$ to this cross section is also significant due to, e.g., 
the zigzag distortion of Ce chains an $a$-direction.     

\section{Low-temperature specific heat of CeNiSn}

To be sure that our approach gives really universal description of 
low-energy spin excitations in CeNiSn, we calculated the magnetic contribuion 
to specific heat of CeNiSn within the approximate mean-field approach to  
RVB thermodynamics with the use of the same parameters for 
the spin fermion spectrum which were chosen for description of the  
neutron scattering data.
Since the samlpes studied in the neutron scattering measurements, 
apparently, were not as perfect as the recently grown samples which 
demonstrate the metallic type conductivity and low value of 
$\gamma=C/T\sim 40$mJ/mole K$^2$ \cite{Taka96}, we used for comparison 
the data for the sample $\# 2$ from \cite{Taka96}. The value of 
$\gamma(T=0.1\;K)=80$mJ/mole K$^2$ was chosen to eliminate the 
ultralow-T upturn due to nuclear and impurity contribution to the 
magnetic specific heat.  
The results of our calculations are presented in fig. 12 (solid 
curve). The experimental data 
presented by diamonds in figure 12 are obtained from those for that sample 
by cancelling the phonon contribution to specific heat. The latter was 
considered to be the same as in LaNiSn crystal \cite{Nolt94}. 

When comparing the experimental 
and theoretical curves for the specific heat, one should take into account 
the contribution of conduction electrons and add it to the  magnetic specific 
heat given by the spin fermion excitations. We have taken it to be linear 
in T within the low-T interval $0<T<10$ K with the Sommerfeld coefficient 
$\gamma_e$=18 mJ/mol K$^2$. This means that the electron contribution to 
the heavy fermion density of states in good samples can be as large as 
$\approx 50$ per cent in agreement with the theoretical predictions for the 
spin liquid state of the Kondo lattices \cite{Kik96}. 
It is seen from figure 12, that the theoretical curve describes quite well 
the temperature behavior of $C/T=\gamma + bT$ at low temperture and gives 
the maximum of $C/T$ at $T=6$ K. The latter is observed for the more perfect 
samples \cite{Nolt95} where the data for higher temperatures are 
available. 

It should be noted that the contribution of spinon dispersion in $a$-direction 
(see Section 5.4) practically does not change the form of $C(T)$ curve since
the van Hove peaks survive in 3D spectrum, and the spinon chemical potential 
is still in the dip of the DOS. 

Our previous semi-phenomenological studies 
have shown that the model of heavy fermion liquid 
with the pseudogap in the 
spin-fermion excitations induced by the interplay with the CF excitations 
describes quantitatively the low-temperature specific heat and thermal 
expansion of CeNiSn  
\cite{KKP93,KVBT94}. Now we see that the consequent microscopic theory gives 
the detailed picture of the low-energy excitations and low-temperature 
thermodynamics of this system which is in  reasonable quantitative agreement 
with the experiment. 
\section{Concluding remarks} 

To conclude, we have found that 
the general theory of spin-fermions acquires some specific 
features in the low-symmetry Kondo lattice  materials with soft CF 
excitations, such as CeNiSn and CeRhSb. These CF excitations modify the 
spectrum of spin excitations in a very characteristic manner, and as a result, 
the density of  spin-fermion states has several peaks due to interaction 
between the spin-fermion excitations and CF excitations. Because of existence 
of these peaks the spin response of CeNiSn and CeRhSb reminds that of 
spin-gap materials.

Whereas the largest peak in spinon DOS at $\approx 2$ meV above the spinon 
Fermi level appears as a result of interplay between 
the HF and CF excitations, all other singularities in one-spinon and 
two-spinon DOS arise  due to quasi two-dimensionality of spin excitation 
spectrum (although these singularities are robust enough against the 
appearance of the third component of spinon dispersion). 
The source of this anisotropy of the excitation spectrum is the 
anisotropy of RKKY interaction in a real space. It is worth mentioning that 
the standard slave-boson description used in a previous phenomenological 
model \cite{KKP93,KVBT94} also predicts highly anisotropic spectrum of HF 
excitations due to the sf-hybridization anisotropy in ${\bf k}$-space. 

If the levels $|\mp 5/2\rangle$ dominate in the ground state Kramers doublet 
$\left|G\pm \right\rangle$ (\ref{2.1a}) (see Section 5.2), 
then the effective intersite overlap 
integral ${\cal T}_{\bf ij}^{GG}$
is determined in this theory 
by the combination of hybridization matrix elements,
$V_{{\bf k}\Lambda}^{{\bf i}*}V_{{\bf k}\Lambda'}^{\bf j}/(\epsilon_F-E_f)$,
which turns into zero for ${\bf k}=[1.0.0]$ (see, e.g., \cite{Hyy87}). This  
means that the f-electron wave functions form the non-bonding states along 
this direction. Moreover, the expression  
$\bar{V}_{kG}^{{\bf i}E*}V_{kG}^{\bf j}/\epsilon_F-E_f)$
which determines in this case the mixing integrals 
${\cal B}_{\bf ij}^{GE}$
turns into zero for ${\bf k}=[1,0,0]$ due to the similar reasons. 
Thus one can expect that the main characteristic features of the interplay 
between the HF and CF states, including the pseudogap, 
would develope in the $bc$-plane of the Brilloun zone. 
However, this kind of theory has an undesirable consequence: in the mean-field 
slave-boson approximation the spin  and charge pseudogaps simply coinside, 
and this statement obviously contradicts the experimental data for CeNiSn. 

Another version of the mean-field description of pseudogap in CeNiSn 
was presented recently in \cite{Ikem96}. It was postulated in this theory that 
the ground state of Ce ion is the doublet $|\mp 3/2\rangle$, and the CF 
excitations where not taken into account. The pseudogap in the HF spectrum 
appeared due to the mentioned non-bonding character of f-f overlap matrix 
element ${\cal T}^{GG}({\bf k})$. Then the low-temperature behavior of 
specific heat and magnetic susceptibility are determined by this pseudogap, 
and the satisfactory agreement between the theoretical description of these 
quantities and the experimental data for CeiSn can be achieved. 

As to the 
inelastic neutron scattering cross section, the theoretical results presented 
in \cite{Ikem96} are too scanty to conclude about agreement 
between the theory and the experiment. One can definitely mention the 
qualitative agreement, at least for the constant-E scans (figure 11 
of Ref.\cite{Ikem96} which 
is more or less similar to our figure 8). As to the constant-${\bf Q}$ scans, 
the slave-boson description is able to describe the pseudogap in the 
two-particle HF DOS (figure 10 of Ref.\cite{Ikem96}), 
although the experimental picture is, 
apparently, much more reach then the smooth curves presented in that figure. 
One should mention also, that the theoretical spectrum for  
${\bf Q}=[1/2,0,0]$ presented in figure 10b demonstrates the peak in  
a-direction which is absent in experiment because of the geometrical factor 
entering the neutron scattering cross section for this direction 
\cite{Mason92,Sato95}. 
It is unclear also whether this theory is able to describe the quasi 
1D character of neutron scattering along the $[0, n+1/2, Q_c]$ direction. 
On the face of it, this task seems to be rather difficult since the matrix
element which governs the HF  dispersion law behaves as 
$V^2(\hat{k}^2_x+\hat{k}^2_y)(1+15\hat{k}^2_z)$ (see equation 2.10b in 
Ref.\cite{Ikem96}). Besides, the CF level scheme adopted in this model is 
not supported by the experimental data available \cite{Aleks94}. 
Finally, the coincidence of the charge and spin pseudogap in the mean field 
slave boson theory mentioned above does not allow to reproduce the metallic
charater of low-temperature resistivity. The absence of $T^2$ in the 
theoretical curve $\rho(T)$ is clearly seen from figure 13 of Ref.\cite{Ikem96} 

The inelastic neutron scattering method  
is an adequate tool for observing the spin-liquid excitations 
in the heavy-fermion 
materials. These measurements give the detailed information concerning the 
density of states of spinon particle-hole excitation spectrum, 
and the dispersion of various branches of spinon excitations can be restored 
by analysing various constant-${\bf Q}$ and constant-$E$ scans. The general 
picture of low-energy excitation spectrum obtained from this analysis 
allowed us to describe also the low-temperature specific heat of CeNiSn
(see also \cite{KKM96}). 
All other important characteristics of this material, 
such as magnetic susceptibility, thermal expansion, NMR relaxation rate, etc, 
also can be described quantitatively within the same  approach. 

The specific features of spin-liquid excitation spectrum 
do not violate the conduction electron 
spectrum which remains essentially metallic, although the an\-iso\-tro\-py 
of spin 
excitations can result in anisotropy of scattering mechanism for conduction 
electrons, and the soft crystal field excitations can influence their 
paramagnetic response. 
In any case, CeNiSn and related materials should be treated as 
metallic Kondo lattices with specific type of spin excitations. 

\section{Acknowledgements}
We are indebted to H. Kadowaki, S. Kambe, and T. Sato for timely 
providing us with experimental information and  important comments, 
to N.V.Prokof'ev for valuable critical remarks  and to
A.Yu.Yaresko for assistance in numerical computations. 
The support of the grants NWO 5-16501, INTAS 93-2834 and RFBR-95-02-04250 
is acknowledged.

\newpage

\appendix
\section{Appendix}

To calculate the anisotropy factor in Cornut-Coqblin Hamiltonian
one should expand the Bloch waves entering the RKKY 
interaction (\ref{1.9}),(\ref{1.10}) in partial waves entering the 
hybridization integrals in equation (\ref{1.4}) and then superimpose 
the direction connecting two ions with the quantization axis by means
of the matrices of finite rotations. 
\par
In the case of short enough 
interionic distances one can neglect the influence of angular dependence of 
the partial waves on the values of radial overlap integrals. This approximation 
\cite{Coqb69}. 
in a simplest case of the doublet formed by the $j=5/2$ states with the same 
$j_z$ value, i.e., $|\Gamma \nu \rangle = |j\; \pm\! M \rangle$  
results in the following equation for the factor of anisotropy of 
exchange interaction, 
$$
B_{G}(\theta)=\left ( B_{M}(\theta) \right)^2 \;,
$$
$$
B_{M}(\theta)=\sum_{l=0,2,4,6}
(-1)^{l/2}(2l+1)Z_l^M P_l(\cos \theta)\;,
$$
$$
Z_l^M=\xi^2_M O(l, M+1/2)+ \eta^2_M O(l, M-1/2)\;, 
$$
\begin{equation}
\xi_M=\sqrt{(7+2M)/14},\;\;
\eta_M=\sqrt{(7-2M)/14},\;\;O(l,j)=C_{30l0}^{30}C_{3jl0}^{3j}
\label{a.1}
\end{equation}
Here $C_{3Mlm}^{3M'}$ are the Clebsch-Gordan coefficients. This 
equation can be found in \cite{Coqb69}  for the case of $\theta=0$.  
Since the CF wave functions transform in accordance with the spherical group 
the only source of anizotropy is the definite direction of quantatization 
axis which lowers symmetry to the axial one. Therefore the interaction 
depends on the polar angle $\theta$ only.
\par
More complicate is the anisotropy factor for a mixed representations 
of the type of eq. (\ref{2.1a}). In this case the wave functions transform 
according to the irreducible representation of the trigonal group with 
third order rotational axis parallel to the quantization axis. In this case 
the RKKY intraction becomes $\phi$-dependent, 
\begin{eqnarray}
B_{G}(\theta,\phi) & = &
\left\{
1 +
\frac{2}{7} \left ( 5 - 9 a^2 \right) P_2(\cos \theta) +
\frac{3}{7} \left ( 1 + a^2  \right) P_4(\cos \theta) \right. \nonumber \\
&&
\left. - \frac{a\sqrt{1-a^2}}{7\sqrt{10}} P^3_4(\cos \theta) cos(3\phi)
\right\}^2
\label{a.2}
\end{eqnarray}
For the specific cases of interaction "in plane" ($\theta=\pi/2$) and along
the quantization axis ($\theta=0$) the factor $B$ does not depend on the 
angle $\phi$. Therefore the ratio between the "in plane" interaction and 
interaction along z-axis can be expressed  in terms of single 
parameter $a$ entering the wave function (\ref{2.1a}), 
\begin{equation}
\frac{B_{G}(\theta=\pi/2)}{B_{G}(\theta=0)}  = 
\left\{
\frac{1}{40} \frac{25+81a^2}{4-3a^2} 
\right\}^2
\label{a.3a}
\end{equation}
It is seen from (\ref{a.3a}) that 
$B_{G}(\theta=\pi/2)/B_{G}(\theta=0) \gg 1$ only when 
$|1/2\rangle$-compomemt of the crystal field wave function (\ref{2.1a})
is dominant.
\par
In the case of the large interionic distances the asymptotic behavior 
of RKKY exchange integrals is dominated by  
$j_z=\pm 1/2$ partial waves  \cite{coop85}, and the "in plane" 
interaction is stronger than that along the quantization axis ($\theta=0$)
in the case when the $|5/2\rangle$-component of the crystal field wave 
function (\ref{2.1a}) is dominant.

\newpage
\begin{center}
{\large\bf Figure captions}
\end{center}

\begin{description}
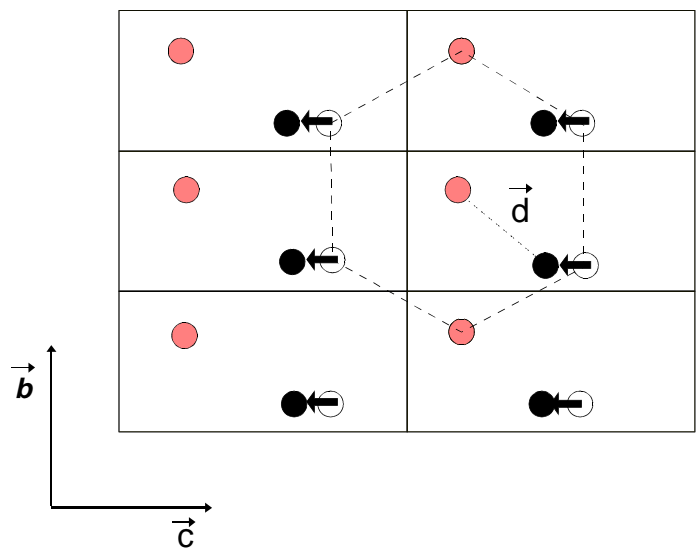
\item[Figure 1:] 
The $bc$ plain of CeNiSn lattice. Two Ce sublattices are
denoted by black and grey circles, resp. 
The orthorhombic distortion {\cal O} (solid
arrows) transforms the simple hexagonal lattice into two-sublattice 
orthorhombic one. The vector {\bf d} is the basis vector of two-ion
elementary cell
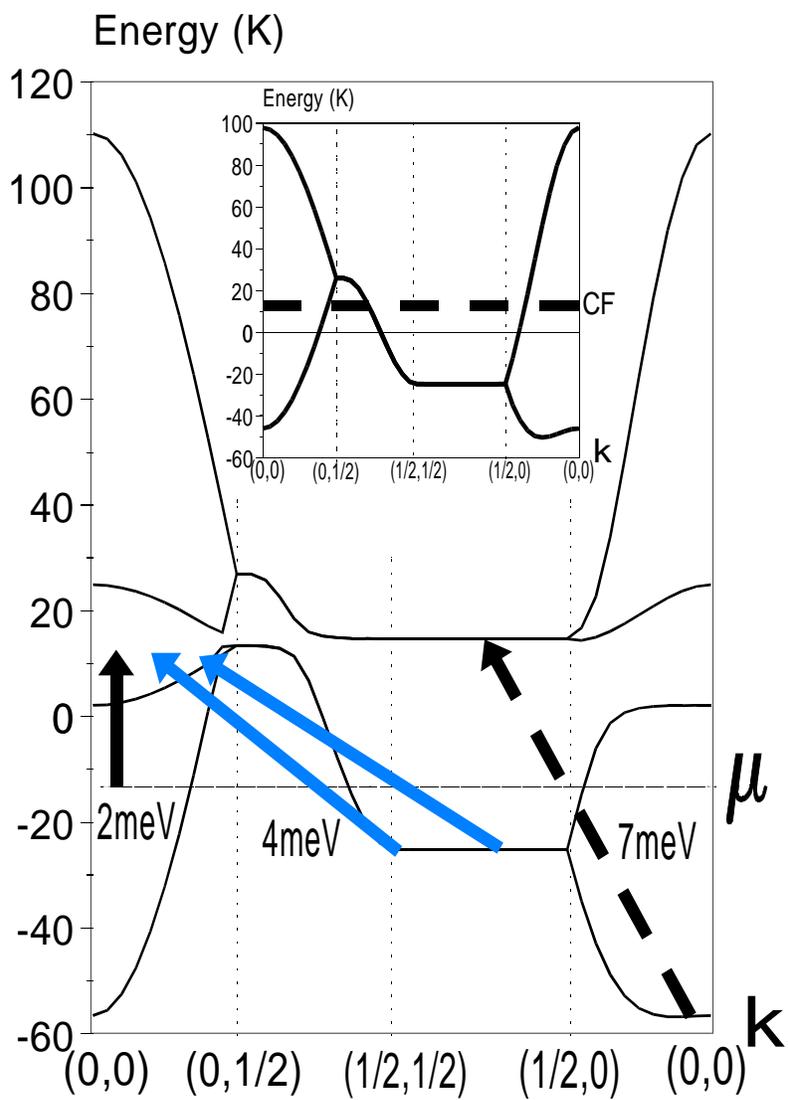
\item[Figure 2:]
The spinon dispersion in CeNiSn along the high symmetry 
directions of 2D Brillouin zone obtained for the following paramenter
values (in Kelvins) of 
the secular matrix (\ref{2.8}): ${\cal T}_1=18.0, {\cal T}_2=12.7, 
\widetilde{\Delta}_{CF}=13.5, {\cal G}_1=3.5, {\cal G}_2=7.7$. 
Horizontal line at the energy $E=-13$ K is the Fermi level position 
at $T=0$.  The arrows indicate the intraband and interband transitions 
responsible for the main peaks in neutron scattering cross section  
(see text).\\
Insert: the non-hybridized spinon spectrum with CF level  
(dased line) superimposed 
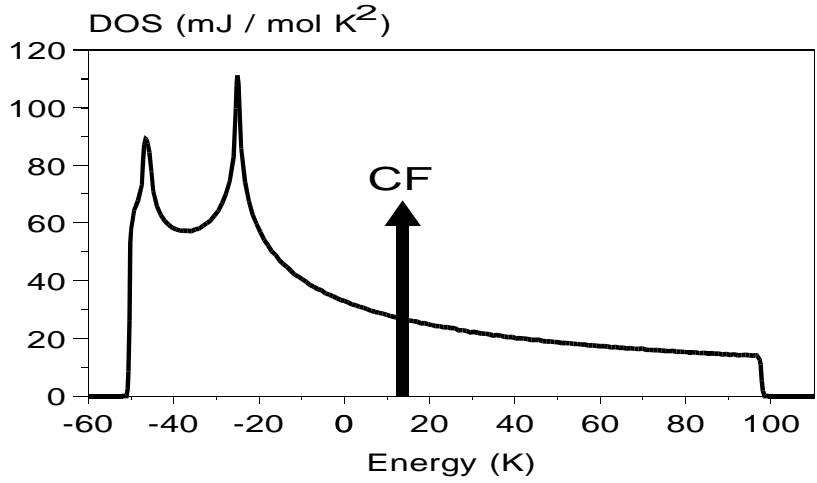
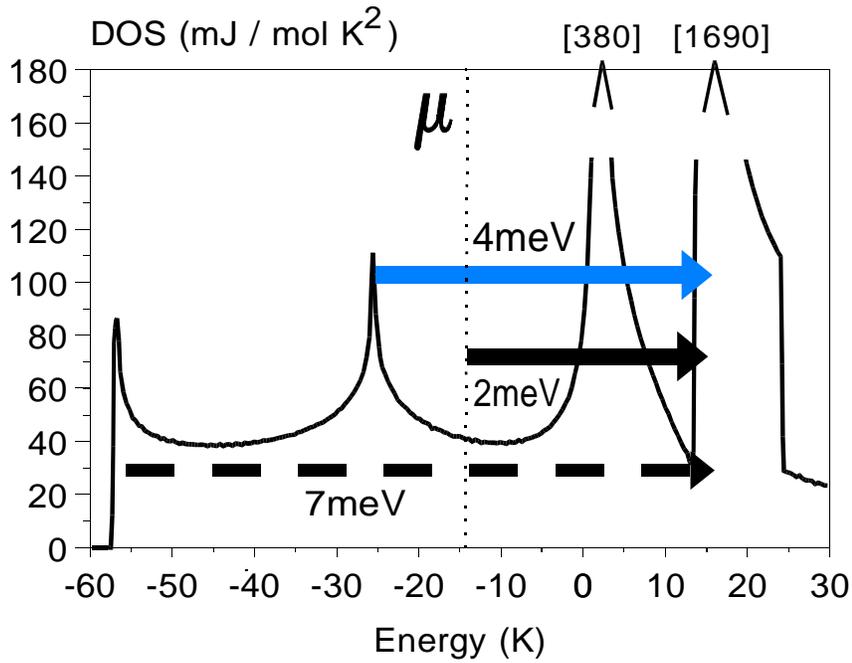
\item[Figure 3:]
Upper panel: the density of states of unhybridized spinon spectrum  \\
Lower panel: the density of states for spin-fermions hybridized with CF 
excitation. All other designations are the same as in fig. 2 
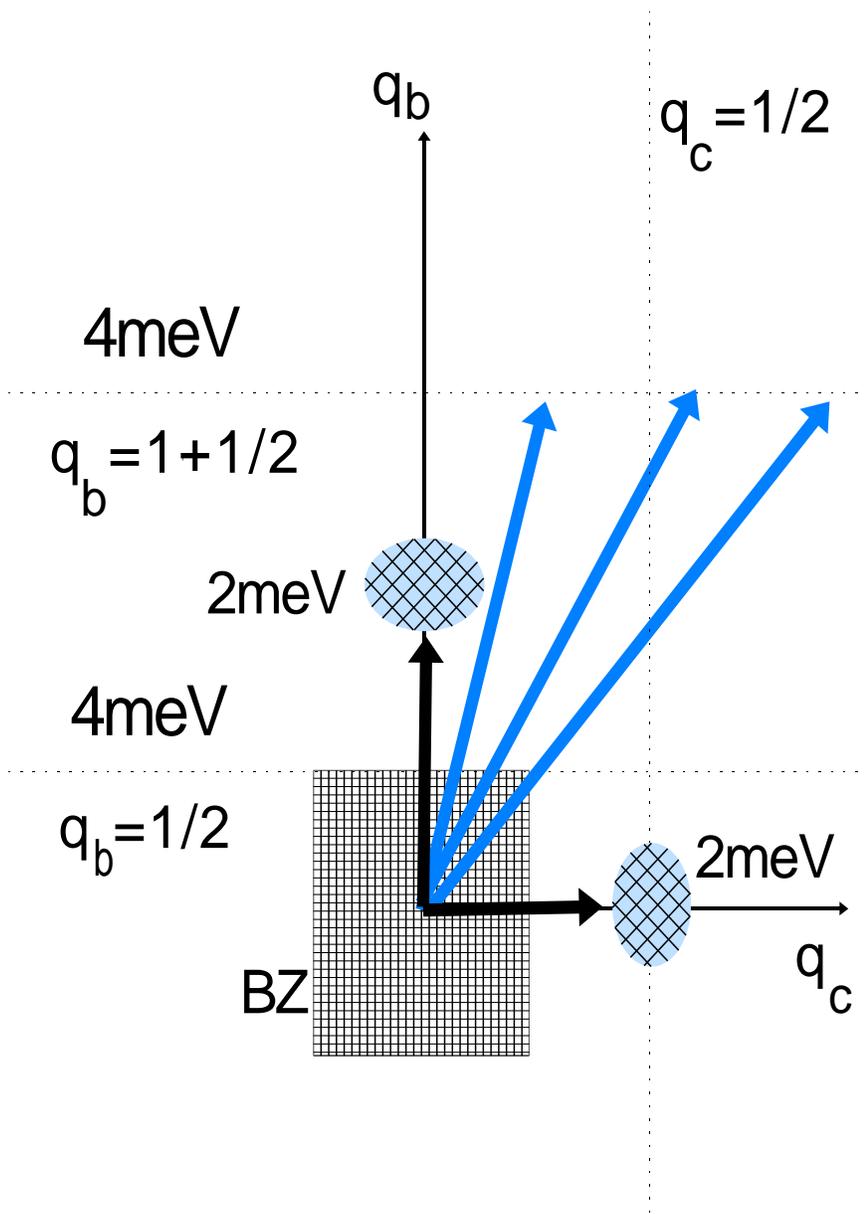
\item[Figure 4:] 
The {\bf Q} vectors  corresponding to the main transitions 
in spinon bands in 2D reciprocal lattice.  
The shaded areas are the projection of BZ onto 
$bc$ plain and the regions in BZ where the final states are located
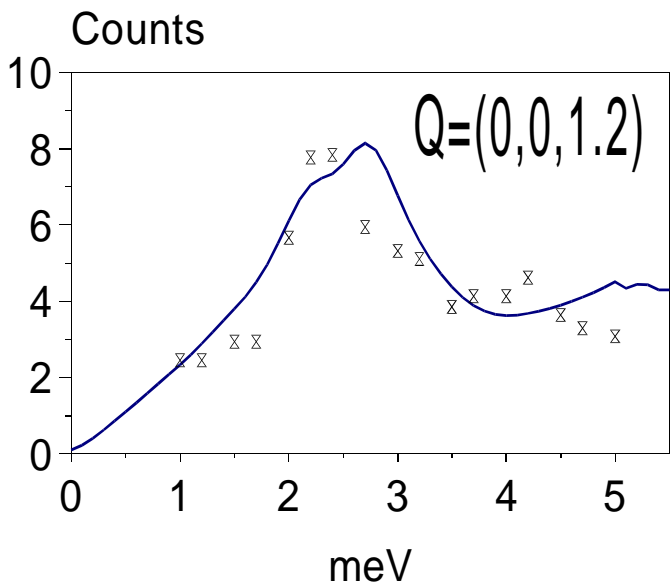
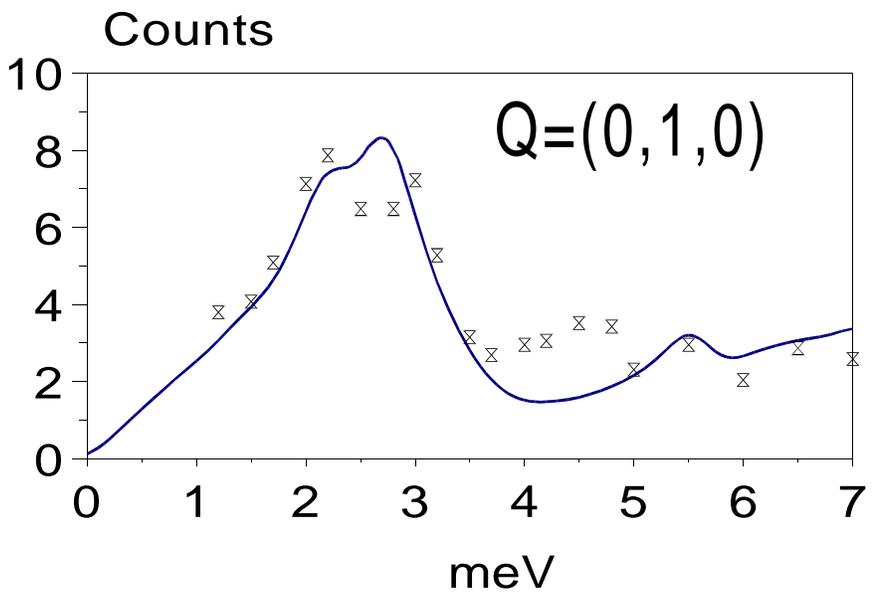
\item[Figure 5:] 
Comparison of experimantal (diamonds) and theoretical (curves) 
constant-{\bf Q} scans. \\
Upper panel: {\bf Q}=(0,0,1.2). Experimental poins are taken from \cite{Mason92}. \\
Lower panel: {\bf Q}=(0,1,0). Experimental points are taken 
from \cite{Kadow94,Sato95}
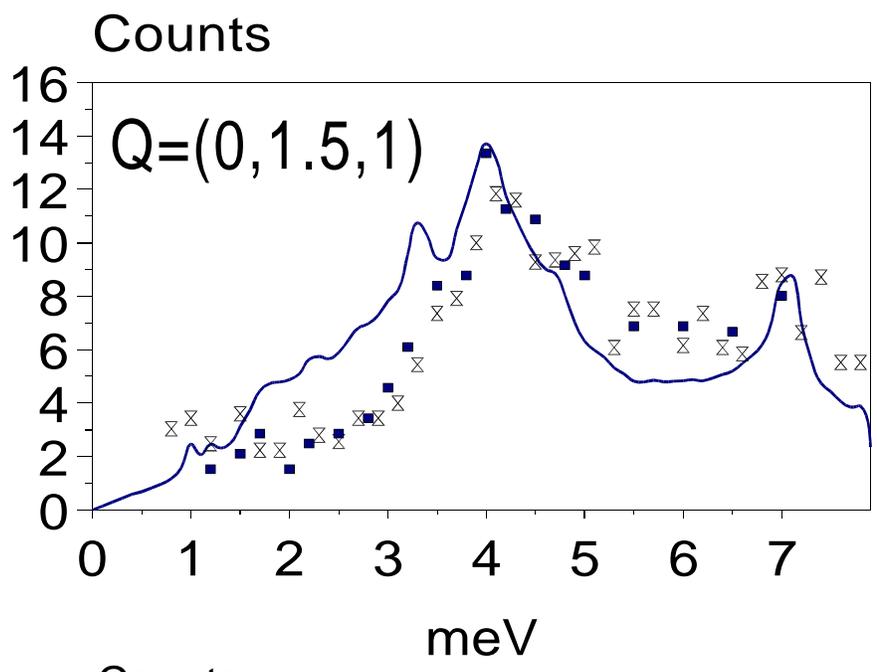
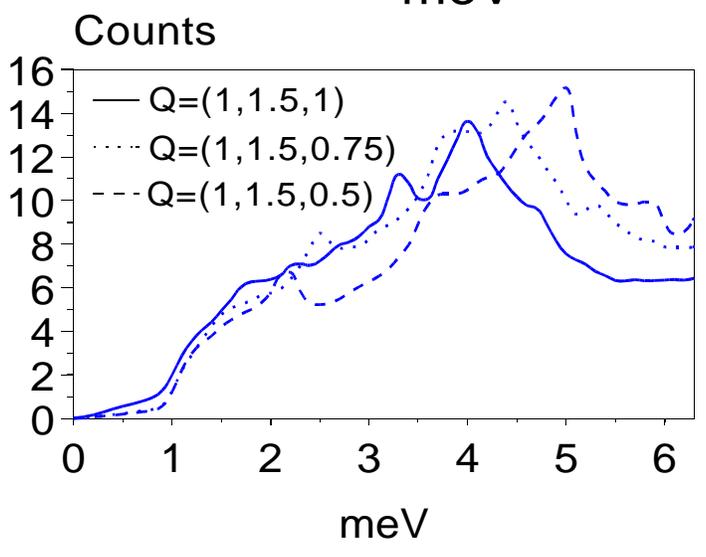
\item[Figure 6:]
Theoretical (lines) and experimental (points) 
constant-{\bf Q} scans with 4meV peak.
Upper panel: Constant-{\bf Q} scan for ${\bf Q}= (0,1.5,1)$. 
Experimental points are taken from \cite{Kadow94,Sato95} (squares) 
and \cite{Kambe96} (diamonds). \\
Lower panel: Constant-{\bf Q} scans for ${\bf Q}= (0,1.5,1)$ (solid line),
${\bf Q}= (0,1.5,0.75)$ (dotted line) and ${\bf Q}= (0,1.5,0.5)$ (dashed line)
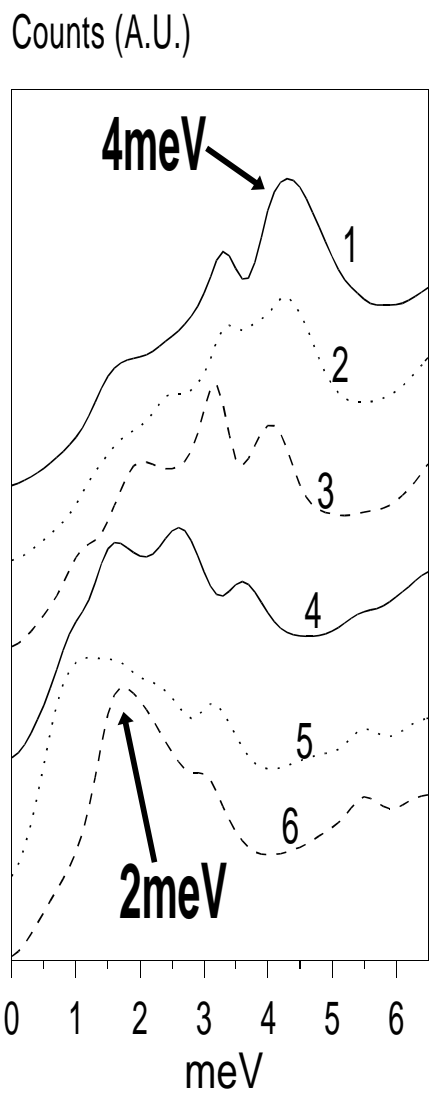
\item[Figure 7:]
Constant-{\bf Q} scans for: {\bf Q}=(0,1.5,0); (0,1.4,0); 
(0,1.3,0); (0,1.2,0); (0,1.1,0); (0,1,0) (curves 1 --6, respectively). 
Position of 4meV and 2meV peaks are indicated by arrows. The vertical scales  
are shifted for clarity 
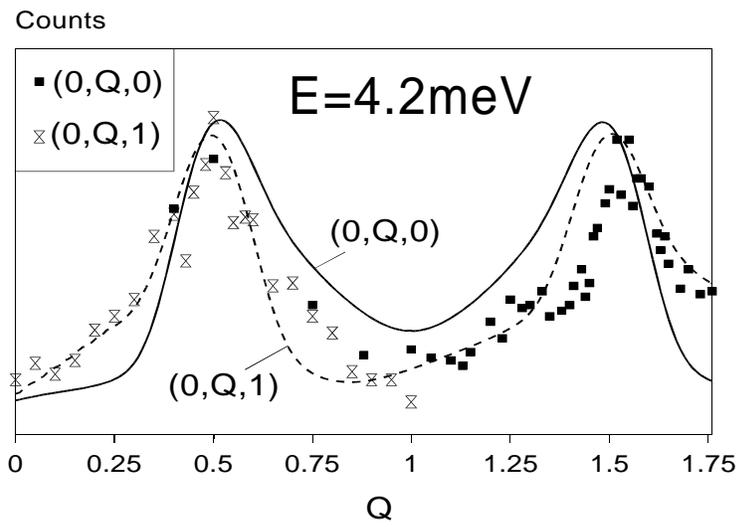
\item[Figure 8:] 
Theoretical (curves) and experimental (dots) 
constant-E scans ($E=4.2$meV) along ${\bf Q}=(0,Q_b,0)$ (squares and solid line) 
and ${\bf Q} = (0,Q_b,1)$ 
(diamonds and dashed line) directions. 
Experimental points are taken from \cite{Sato95}
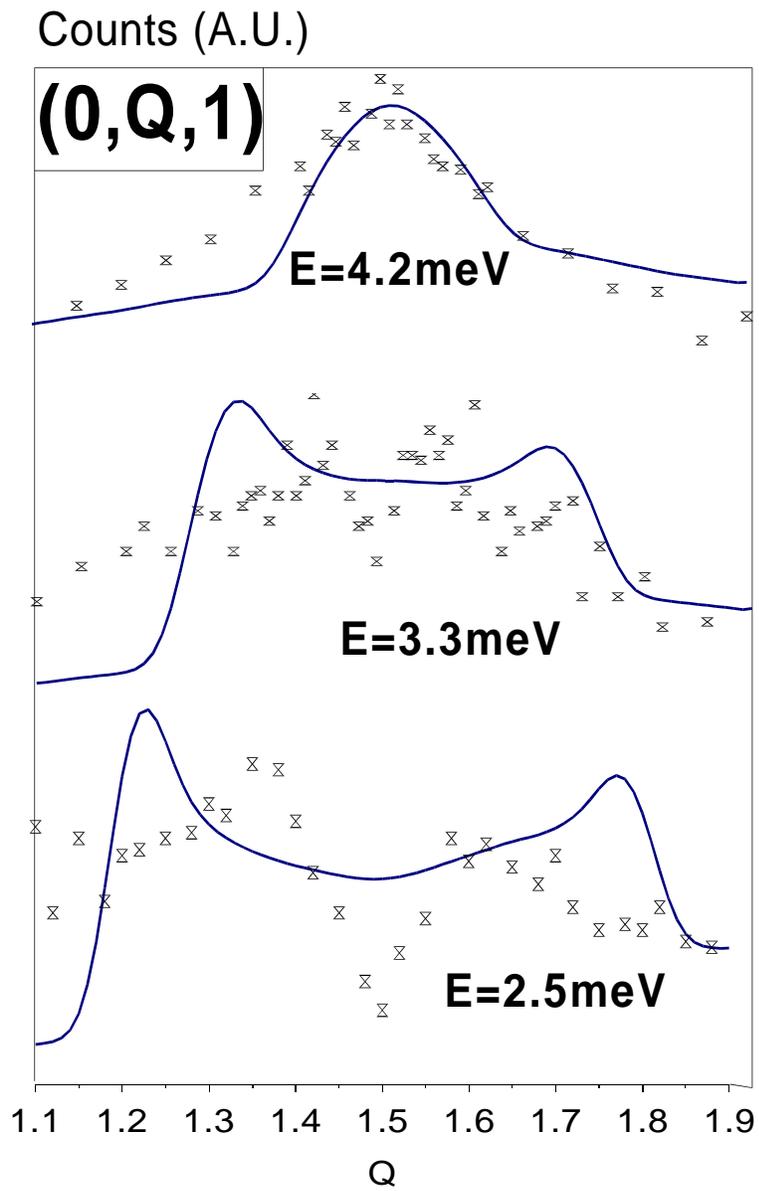
\item[Figure 9:]
Theoretical (curves) and experimental (dots) 
constant-E scans along $(0,Q_b,1)$ direction: (a) E=4.2mV; (b) E=3.3meV; (c) E=2.5meV.
Experimental points are taken from \cite{Kambe96} 
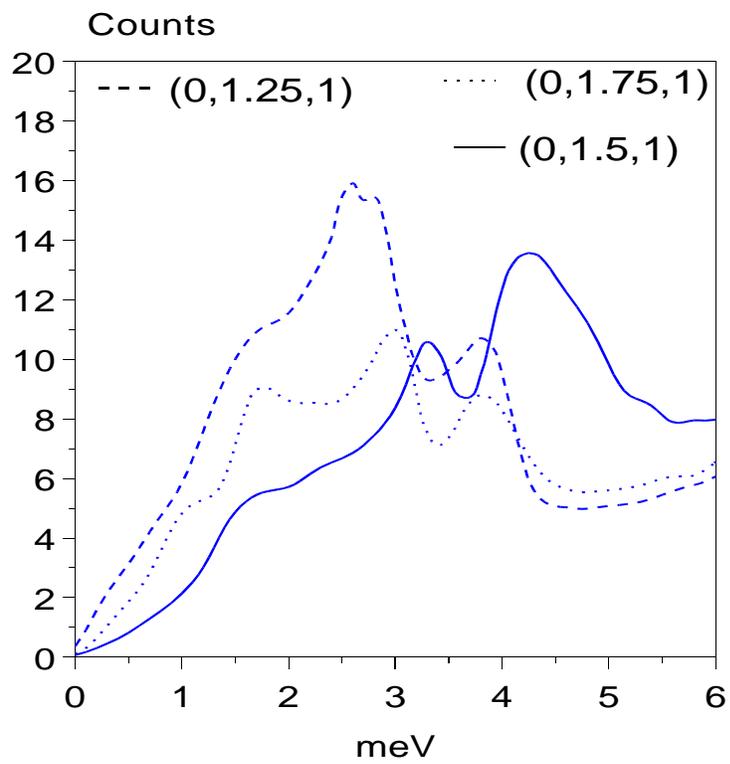
\item[Figure 10:]
Constant-{\bf Q} scans for: ${\bf Q}=(0,1.25,1)$ (dashed line); 
${\bf Q}=(0,1.5,1)$ (solid line); ${\bf Q}=(0,1.75,1)$ (dotted line).
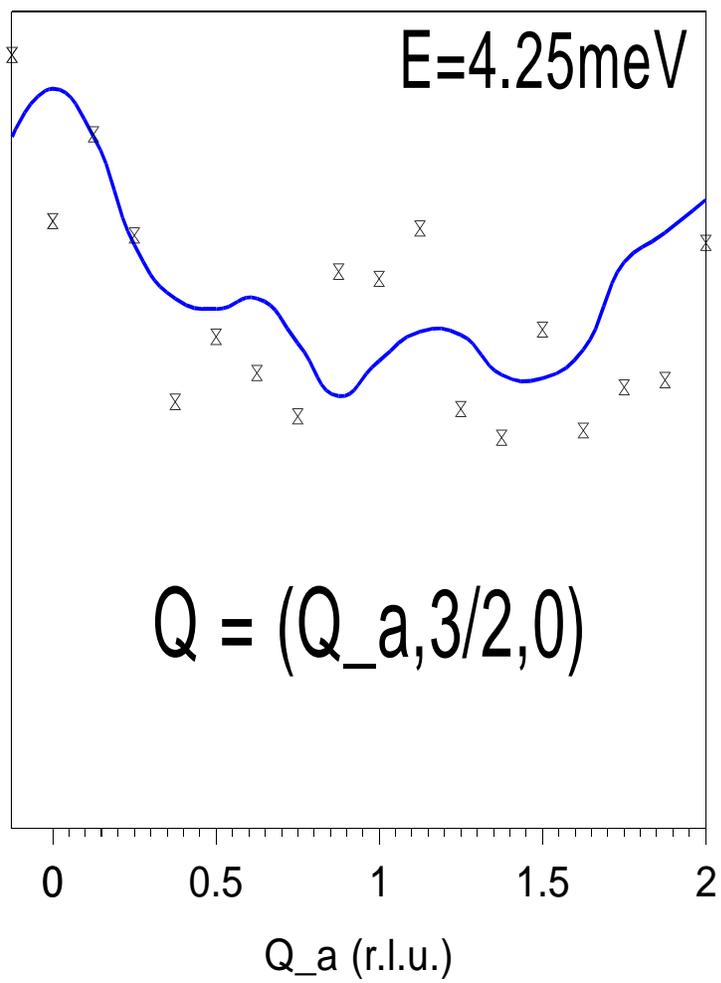
\item[Figure 11:]
Theoretical (curves) and experimental (dots) 
constant-E scans ($E=4.25$meV) along ${\bf Q}=(Q_a,3/2,0)$ 
direction. 
Experimental points are taken from \cite{Sato95}
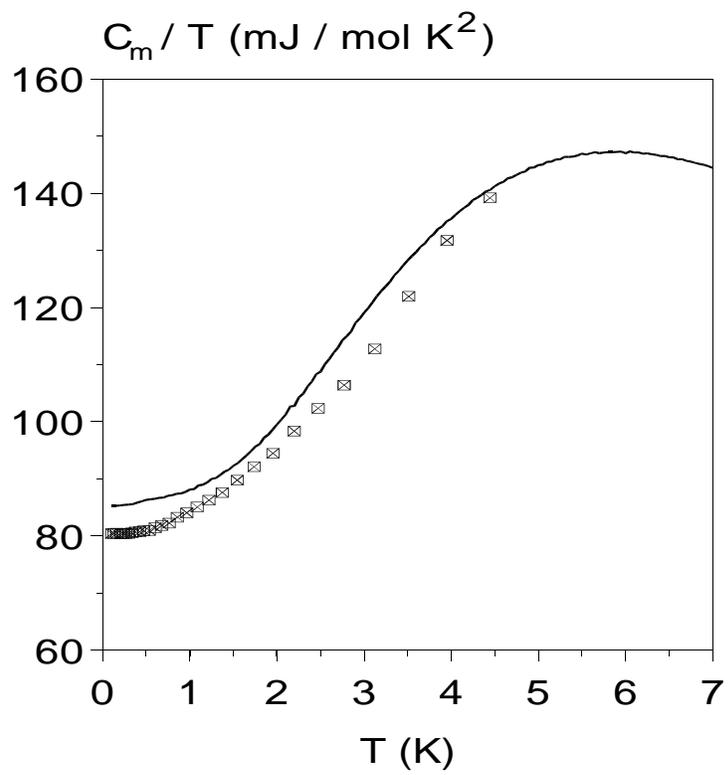
\item[Figure 12:]
Theoretical curve for low-T contribution to the specific heat 
in comparison with  the experimental data taken from \cite{Taka96} 
(diamonds). The constant electronic contribution $\gamma_{el}=18$ mJ/mol K$^2$ 
is added to the theoretical curve (see text) 
\end{description}

\end{document}